\documentclass[12pt]{article} 
\usepackage{amssymb}
\usepackage{amsmath} 
\usepackage{cite}
\usepackage{axodraw}
\usepackage{hyperref}
\input xy
\xyoption{all}
\usepackage{rotfloat}   

\def\sqr#1#2{{\vcenter{\vbox{\hrule height.#2pt
            \hbox{\vrule width.#2pt height#1pt \kern#1pt
                  \vrule width.#2pt}\hrule height.#2pt}}}}

\def\square
 {\mathop{\mathchoice{\sqr{12}{15}}{\sqr{9}{12}}
{\sqr{6.3}{9}}{\sqr{4.5}{9}}}}

\def\sqra#1#2#3{{\vcenter{\vbox{\hrule height.#2pt
            \hbox{\vrule width.#2pt height#1pt \kern5pt 
#3
                  \vrule width.#2pt}\hrule height.#2pt}}}}

\numberwithin{equation}{section}

\numberwithin{table}{section}

\begin{document} 

\begin{center}

{\large\bf Dilaton shifts, probability measures,
and decomposition}

\vspace*{0.2in}

Eric Sharpe

Department of Physics MC 0435\\
850 West Campus Drive\\
Virginia Tech\\
Blacksburg, VA  24061

{\tt ersharpe@vt.edu}

\end{center}

In this paper we discuss dilaton shifts (Euler counterterms)
arising in decomposition of two-dimensional
quantum field theories with higher-form symmetries.
These take a universal form, reflecting underlying (noninvertible, quantum)
symmetries.
The first part of this paper constructs
a general formula for such dilaton shifts,
and discusses related computations.
In the second part of this paper,
we comment on the relation between decomposition and ensembles.

\begin{flushleft}
December 2023
\end{flushleft}

\newpage

\tableofcontents

\newpage

\section{Introduction}

Put simply, for a local quantum field theory to decompose means that 
it
is equivalent to a disjoint union of other
local quantum field theories (known in this context as `universes').  
A quantum field theory that one thought one knew, might secretly be
a union of several independent quantum field theories masquerading as a unit.
Decomposition was first observed in
examples in \cite{Hellerman:2006zs}, where it was used to 
resolve apparent inconsistencies in string compactifications
on certain stacks 
known as gerbes
\cite{Pantev:2005rh,Pantev:2005wj,Pantev:2005zs},
which are
fiber bundles of higher-form symmetry groups, realized physically
as gauge theories with trivially-acting subgroups.

Decomposition has been discussed and applied by now extensively to a variety
of examples, including
Gromov-Witten theory (see e.g.~\cite{ajt1,ajt2,ajt3,tseng1,gt1,xt1}),
gauged linear sigma models
(see e.g.~\cite{Caldararu:2007tc,Hori:2011pd,Halverson:2013eua,Hori:2013gga,Hori:2016txh,Wong:2017cqs,Kapustka:2017jyt,Chen:2020iyo,Guo:2021aqj,Katz:2022lyl,Katz:2023zan,Lee:2023piu}),
elliptic genera and IR limits of pure supersymmetric
gauge theories (see e.g.~\cite{Eager:2020rra}), 
adjoint QCD$_2$ \cite{Komargodski:2020mxz},
anomaly resolution \cite{Robbins:2021lry,Robbins:2021ibx,Robbins:2021xce},
lattice computations \cite{Honda:2021ovk},
and even quivers \cite{Meynet:2022bsg},
in not only two-dimensional theories but
also three-dimensional (see e.g.~\cite{Pantev:2022kpl,Pantev:2022pbf,Perez-Lona:2023llv})
as well as four-dimensional gauge theories
(see e.g.~\cite{Tanizaki:2019rbk,Cherman:2020cvw}).
It is often associated with the existence of
higher-form symmetries -- specifically, a (possibly noninvertible
\cite{Nguyen:2021yld,Nguyen:2021naa,Sharpe:2021srf,Huang:2021zvu}) $(d-1)$-form 
symmetry in a $d$-dimensional quantum field theory 
\cite{Tanizaki:2019rbk,Cherman:2020cvw}.
In such theories, decomposition often has the effect of restricting
allowed instantons, through a ``multiverse interference effect'' created
by the superposition of multiple quantum field theories (the universes).
See e.g.~\cite{Sharpe:2010zz,Sharpe:2010iv,Sharpe:2019yag,Sharpe:2022ene}
for reviews and additional references.

As first observed in \cite{Hellerman:2006zs},
the contributions to a partition function from different universes
often have different Euler number counterterms, universal counterterms multiplying
worldsheet Ricci curvatures, which in a string theory would correspond
to constant dilaton shifts.  For this reason, we refer to these counterterms
as ``dilaton shifts,'' as was the usage in \cite{Hellerman:2006zs}.

The purpose of this paper is to systematically study these dilaton shift
factors arising between the different universes in decomposition in
two dimensions.  These factors have a canonical and universal form,
reflecting symmetries of the decomposing theory, which we will discuss.

These dilaton shifts are also sometimes interpreted in the form
of probability densities.  Now, a decomposition is not the same
as an ensemble, as we shall discuss, but at least on a connected spacetime,
they do appear closely related, which is one motivation to understand
dilaton shift factors more systematically.

We begin in section~\ref{sect:decomprev} with a brief review of
decomposition in two-dimensional theories.  
In section~\ref{sect:dilaton-shift} we give general results for
dilaton shifts in orbifolds, gauge theories, and topological field
theories in two dimensions, which we check in numerous examples.
Briefly, the dilaton shifts are universally proportional to
$(\dim R)^2$, where for orbifolds and gauge theories,
$R$ is an irreducible representation of some gauge group indexing
the universes, and in topological field theories, a ratio of
$\dim R$'s is the quantum dimension of an interface between universes.
As an aside, in section~\ref{sect:exs:gauge:volumes}, we discuss the
implications of decomposition for volumes of moduli spaces of flat
connections.

In section~\ref{sect:consistency-asymp} we discuss how this form is
expected on symmetry grounds, related to the presence and properties of
interfaces linking the different universes, and also relate these shifts
to asymptotic densities of states as in e.g.~\cite{Kapec:2019ecr}.
In section~\ref{sect:tftcoupling} we outline how the form of this
result for dilaton shifts can also be understood by reinterpreting at least
some of these theories in the language of coupling to a topological field
theory.

In section~\ref{sect:ds-vs-prob} we compare and contrast these
dilaton shifts with probability measures appearing in various
discussions.  In particular, a decomposition is not the same as
an ensemble, as we discuss, though in the special case of
a connected spacetime,
they are at least naively closely related.
We compare structures in mirrors to decompositions to
stochastic variables appearing in the SYK construction,
and compute entanglement entropy in decomposing theories.

In appendix~\ref{app:inv} we briefly review the notion of invertible
field theories, which arise in numerous locations in this paper.
In appendix~\ref{app:reptheory} we collect some finite group 
representation theory identities, which are used in a number of places.
In appendix~\ref{app:2ddw:triangle} we give a presentation of 
two-dimensional Dijkgraaf-Witten theory utilizing triangulations.

\section{Brief review of decomposition}
\label{sect:decomprev}

Briefly, decomposition is a property of local quantum field theories,
in which they are equivalent to (``decompose into'') a disjoint union
of other local quantum field theories, known in this context as
`universes.'
Formally, decomposition is expected to arise in any
$d$-dimensional quantum field theory with a global
$(d-1)$-form symmetry (see e.g.~\cite{Sharpe:2022ene} for a recent review).

One of the signatures of decomposition in a unitary quantum field
theory is the existence of
topological projection operators in the spectrum of local operators,
a set of operators $\Pi_i$ that commute with all other operators,
\begin{equation}
[ \Pi_i, {\cal O} ] \: = \: 0,
\end{equation}
and which behave like projectors in the sense that
\begin{equation}
\Pi_i \Pi_j \: = \: \delta_{ij} \Pi_i, \: \: \: 
\sum_i \Pi_i \: = \: 1.
\end{equation}
As the projectors $\Pi_i$ commute with local operators, they are
simultaneously diagonalizable, with eigenspaces corresponding to the
states associated to the different universes,
onto which they project.
Another signature of decomposition is that partition functions
on connected worldsheets
can be written as sums, schematically
\begin{equation}
Z \: = \: \sum_i Z_i,
\end{equation}
a consequence of the fact that the state space breaks up into eigenspaces
of the projection operators.
The projection operators are also the conserved defect operators of a
$(d-1)$-form symmetry.

Simple examples of decomposition in two-dimensional theories are
gauge theories with trivially-acting subgroups
\cite{Pantev:2005rh,Pantev:2005wj,Pantev:2005zs,Hellerman:2006zs}.
Suppose, for example, we have a $\Gamma$ gauge theory (where $\Gamma$
might be either finite -- an orbifold -- or continuous -- an ordinary
gauge theory) in two dimensions, in which a subgroup
$K \subset \Gamma$ acts trivially.  For simplicity, let us assume
that $K$ lies within the center of $\Gamma$.  (Decomposition is understood
in more general cases, but for our purposes here, this will suffice.)
Then, schematically,
\begin{equation}  \label{eq:decomp1}
{\rm QFT}\left( \Gamma-\mbox{gauge theory} \right)
\: = \:
\coprod_{\theta \in \hat{K}} {\rm QFT}\left( (\Gamma/K)-\mbox{gauge theory}
\right)_{\omega(\theta)},
\end{equation}
where the $\omega(\theta)$'s indicate discrete theta angles,
determined by the image of the extension class $[\omega] \in 
H^2(\Gamma/K, K)$ corresponding to
\begin{equation}
1 \: \longrightarrow \: K \: \longrightarrow \: \Gamma \: \longrightarrow \:
\Gamma/K \: \longrightarrow \: 1
\end{equation}
under the map 
\begin{equation}
H^2(\Gamma/K, K) \: \longrightarrow \: H^2(\Gamma/K, U(1))
\end{equation}
induced by $\theta: K \rightarrow U(1)$.
(If $\Gamma$ is finite, then the `discrete theta angles' are choices
of discrete torsion.)

Now, to be clear, the expression~(\ref{eq:decomp1}) has glossed over
the `dilaton shifts' (Euler counterterms) that will be the focus of this paper.
Indeed, as these dilaton shifts are merely counterterms, which can be
added or subtracted at will, the expression~(\ref{eq:decomp1}) is still
correct.  However, if we pull them out explicitly into
(universe-dependent) factors $\rho(\theta)$,
then expression~(\ref{eq:decomp1}) can be rewritten (on a connected
worldsheet $\Sigma$) as
\begin{equation}  \label{eq:decomp2}
{\rm QFT}\left( \Gamma-\mbox{gauge theory} \right)
\: = \:
\coprod_{\theta \in \hat{K}} \rho(\theta)^{\chi(\Sigma)} \,
{\rm QFT}\left( (\Gamma/K)-\mbox{gauge theory}
\right)_{\omega(\theta)},
\end{equation}
which is the starting point for the analysis of this paper.

Decomposition is not restricted to gauge theories, but in fact
exists in more general examples.  Important special cases include
two-dimensional unitary topological field theories, with semisimple
local operator algebras.

It is a standard result that two-dimensional unitary topological
field theories with semisimple local operator algebras are
equivalent to disjoint unions of `trivial' (invertible field theories,
see e.g.~\cite{Durhuus:1993cq}, \cite[section 3.1]{Moore:2006dw}, which in
e.g.~\cite[section 3]{Huang:2021zvu},
\cite[appendix C.1]{Komargodski:2020mxz} was observed to be a decomposition.
In broad brushstrokes, such topological field theories have
a decomposition of the form
\begin{equation} \label{eq:common}
\coprod_R {\rm Inv}(0, \ln( \dim R) ),
\end{equation}
up to an overall dilaton counterterm,
where Inv$(\lambda_1,\lambda_2)$ labels a familiy of invertible
field theories, and the decomposition is over irreducible
representations.  
(See appendix~\ref{app:inv} for a review of invertible field theories
and our labelling conventions.)
For Dijkgraaf-Witten, $BF$ theory, and pure Yang-Mills,
these will all be ordinary or projective representations; for
the $G/G$ model, these will be integrable representations, and
dim $R$ the quantum dimension.
(For completeness, some excellent references on the two-dimensional topological
gauge theories appearing in this paper are
\cite{Birmingham:1991ty,Thompson:1992hv,Blau:1991mp,Blau:1993hj}.)

Before going on, let us compare
cohomological topological field
theories
to the Schwarz-type theories
we study in this paper.  Such theories, such as for example the
A and B models in two dimensions, have position-independent (hence topological)
operators, and hence their topological subsectors decompose.
However, no decomposition of the underlying quantum field theory
is expected, unless of course
the target space has multiple
disconnected components or is a gerbe.
One difference is that constructions of
projection operators from dimension-zero operators rest on a physical assumption
of unitarity, which does not typically hold after topological twisting.
For this reason, when we speak of decomposition and dimension-zero
operators, we refer to untwisted theories, and so exclude the
topological A and B models, topological gravity, and topological
Kazama-Suzuki cosets.  (Put another way, the topological subsector
might decompose abstractly as a Frobenius algebra, but one does not
expect the quantum field theory in which the topological sector is embedded
will decompose.)  There can still exist
untwisted topological field theories, and these include
Dijkgraaf-Witten theory, $BF$ theories, and $G/G$ models.

\section{Dilaton shifts in decomposition in two dimensions}
\label{sect:dilaton-shift}

A subtlety of decomposition in two dimensions is that in computing 
partition functions on surfaces of genus $g \neq 1$, there can be a 
dilaton\footnote{
To be clear, these theories are need not be coupled to worldsheet gravity.
We refer to this as a ``dilaton shift'' solely because of the universal
form of the counterterm -- proportional to the worldsheet curvature.
Such counterterms can appear in any two-dimensional theory, regardless
of any worldsheet gravity coupling or lack thereof.
}
shift (Euler counterterm)
appearing on the universes -- a counterterm in the action proportional
to the worldsheet curvature, that generates a genus-dependent factor
multiplying the partition function.

In this section, we conjecture a universal form for such dilaton shifts
in orbifolds, gauge theories, and topological field theories,
namely that they are proportional to $(\dim R)^2$.
In orbifolds and gauge theories, $R$ is an irreducible representation
of some group indexing the universes; in topological field theories,
$\dim R$'s is instead interpreted as the quantum dimension of an 
an operator.  (Other factors vary slightly due to differences
in normalization conventions, so a single result valid for all cases
is not possible, a statement of proportionality is the most one can
hope to expect.)

In section~\ref{sect:orb:nodt} we discuss two-dimensional orbifolds with
trivially-acting subgroups and no discrete torsion.  We formulate a precise
conjecture for the dilaton shift factors, which we check in a number of
examples.  In section~\ref{sect:orb:dt} we repeat for orbifolds with
discrete torsion, making a conjecture which is checked in numerous examples.
In section~\ref{sect:exs:qs} we consider orbifolds with quantum symmetries,
again formulating a conjecture which is checked in examples.

In section~\ref{sect:exs:gauge} we perform the same analysis in two-dimensional
gauge theories in which a subgroup of the gauge group acts trivially.
We also make some general observations about relations between volumes of
moduli
spaces of flat connections implied by decomposition, 
in section~\ref{sect:exs:gauge:volumes}.
Finally, in section~\ref{sect:exs:tfts} we discuss dilaton shifts in
examples of unitary two-dimensional topological field theories..

Later we will discuss general arguments for these dilaton shift factors,
and compare to probability densities.

\subsection{Examples in orbifolds without discrete torsion}
\label{sect:orb:nodt}

\subsubsection{Conjecture}
\label{sect:orb:nodt:conj}

In an orbifold $[X/\Gamma]$ or gauge theory, for $K \subset \Gamma$
acting trivially, 
decomposition predicts that  \cite{Hellerman:2006zs}
\begin{equation}  \label{eq:decomp:nodt}
{\rm QFT}\left( [X/\Gamma] \right) \: = \: {\rm QFT}\left(
\left[ \frac{ X \times \hat{K} }{ \Gamma/K } \right]_{\hat{\omega}} \right)
\end{equation}
where $\hat{\omega}$ denotes discrete torsion, and the right-hand side
in general is a sum of multiple disconnected components (details depending
upon the action of $\Gamma/K$ on the set of irreducible representations
$\hat{K}$ of $K$).  (For example, if $K$ is central in $\Gamma$, then
$\Gamma/K$ acts trivially on $\hat{K}$, and there are as many different
components, as many universes, as elements of $\hat{K}$ (irreducible
representations of $K$)).
The expression above omits dilaton shift (Euler counterterm) factors.
The reader should note that if $\Gamma/K$ exchanges several elements of
$\hat{K}$, then the irreducible representations appearing in that
orbit all have the same dimension.

We conjecture that the genus $g$ partition function
can be written in the following form, which makes dilaton
shifts explicit: 
\begin{equation}  \label{eq:dilaton-shift-conj}
Z_g\left( [X/\Gamma] \right)
 \: = \:  \sum_U \left( \frac{ \dim R_U }{|K|} \right)^{2-2g} Z_g\left( X_U \right), 
\end{equation}
generalizing equation~(\ref{eq:decomp:nodt}) above, 
where $U$ denotes universes,
$\dim R_U$ is the dimension of a representative\footnote{
As opposed to e.g.~the sum of all
representations in the orbit.
} irreducible representation
appearing in the orbit of $G = \Gamma/K$,
and where $X_U$ is the theory corresponding to 
universe $U$.  

In the expression above,
we define the orbifold partition function
without a dilaton shift
by the normalization
\begin{equation}
Z_g\left( [X/\Gamma] \right) \: = \: \frac{1}{|\Gamma|}
 \sum_{a_i, b_j} Z(a_1, \cdots, b_g).
\end{equation}
In some of the papers we cite, partition functions are normalized
differently, with a factor of $1/|\Gamma|^g$ instead of
$1/|\Gamma|$.  We shall denote
such partition functions by $\tilde{Z}_g$, as
\begin{equation}
\tilde{Z}_g\left( [X/\Gamma] \right)
 \: = \: \frac{1}{|\Gamma|^g} \sum_{a_i, b_j} Z(a_1, \cdots, b_g),
\end{equation}
to distinguish them from our conventions above.

The two normalizations above are related by a dilaton shift (a factor of
$1/|\Gamma|^{g-1}$). More generally, we can 
always add an Euler counterterm to the ambient theory,
so that there isn't ``one'' partition function so much as a one-parameter
family. 

Taking into account that ambiguity in shifting the ambient theory by 
Euler counterterms,
a more invariant statement of the conjecture 
is that the relative dilaton shift
between the partition function contribution from a single universe $U$
and that of the ambient theory
is
\begin{equation}
\left( \frac{ \dim R_U }{|K|} \right)^{2-2g}.
\end{equation}
Later in section~\ref{sect:consistency-asymp} we will
discuss how the factors of $\dim R$ seem to be determined by symmetries.

We shall see that this matches dilaton shifts in orbifolds in which $K$
is both abelian and nonabelian.
Later in section~\ref{sect:orb:dt} we will generalize this expression to
orbifolds with discrete torsion.  We will give an expression for dilaton
shift factors which is of an identical form; however, the interpretation
of the $R_U$ will differ slightly, as we elaborate there.

Our conjecture above generalizes the results in
\cite{Kapec:2019ecr} for matrix ensemble eigenvalue densities,
see for example \cite[equ'n (43)]{Kapec:2019ecr}.
We shall return to this in section~\ref{sect:consistency-asymp}.

In passing, these dilaton shift factors have also been studied in
connection with
Frobenius-Schur indicators, see e.g.~\cite{Ichikawa:2020ipg}.

\subsubsection{Examples with trivially-acting 
central subgroups}
\label{sect:ex:trivcentral}

In this section we look at orbifolds 
in which the trivially-acting
subgroup lies in the center, beginning with $[X/D_4]$ with
trivially-acting central ${\mathbb Z}_2$.

From\footnote{
The reader should note that the formula for the $g$ loop partition function
of this orbifold in \cite[section 5.2]{Hellerman:2006zs} has a typo:
the $|D_4|^2$ should instead be $|D_4|^g$.
} \cite[section 5.2]{Hellerman:2006zs},
on a genus $g$ Riemann surface,
it was shown that
\begin{equation}
\tilde{Z}_g\left( [X/D_4] \right) \: = \: 2^{g-1}\left(
\tilde{Z}_g\left( [X/{\mathbb Z}_2 \times {\mathbb Z}_2] \right) \: + \:
\tilde{Z}_g\left( [X/{\mathbb Z}_2 \times {\mathbb Z}_2]_{\rm dt} \right)
\right),
\end{equation}
in conventions in which the partition function is normalized
by $1/|\Gamma|^g$.  Converting to our conventions with
\begin{equation}
\tilde{Z}_g\left( [X/D_4] \right) \: = \: 
\frac{1}{|D_4|^{g-1}} Z_g\left( [X/D_4] \right), 
\: \: \:
\tilde{Z}_g\left( [X/{\mathbb Z}_2 \times {\mathbb Z}_2] \right) \: = \:
\frac{1}{| {\mathbb Z}_2 \times {\mathbb Z}_2|^{g-1}}
Z_g\left(  [X/{\mathbb Z}_2 \times {\mathbb Z}_2] \right),
\end{equation}
we find
\begin{equation}
Z_g\left( [X/D_4] \right) \: = \: 2^{2g-2}\left(
Z_g\left( [X/{\mathbb Z}_2 \times {\mathbb Z}_2] \right) \: + \:
Z_g\left( [X/{\mathbb Z}_2 \times {\mathbb Z}_2]_{\rm dt} \right)
\right).
\end{equation}

Each of the universes (each a ${\mathbb Z}_2 \times
{\mathbb Z}_2$ orbifold) is associated with a one-dimensional
irreducible representation of $K = {\mathbb Z}_2$,
and as $|K| = 2$, we see that the result above matches the
general prediction~(\ref{eq:dilaton-shift-conj}).

More generally, for
\begin{equation}
1 \: \longrightarrow \: K \: \stackrel{\iota}{\longrightarrow} \: \Gamma \:
 \stackrel{\pi}{\longrightarrow} \:
G \: \longrightarrow \: 1,
\end{equation}
for $K$ finite, central, and trivially-acting,
the same analysis as \cite[section 5]{Hellerman:2006zs} yields
\begin{equation}
Z_g\left( [X/\Gamma] \right) \: = \:
\frac{ |G| }{ |\Gamma| } |K|^{2g} \left(
|K|^{-1}
\sum_{\rho \in \hat{K}} Z_g\left( [X/G]_{\hat{\omega}(\rho)} \right)
\right).
\end{equation}
Since
\begin{equation}
\frac{ |G| }{ |\Gamma| } |K|^{2g-1} \: = \: |K|^{2g-2},
\end{equation}
and $K$ is central, the representation $R_U$ of $K$ associated to
each universe is one-dimensional.  Thus, the partition function above
can be written in the form
\begin{equation}
Z_g\left( [X/\Gamma] \right) \: = \:
|K|^{2g-2} \sum_{\rho \in \hat{K}} (1)^{2-2g}  Z_g\left( [X/G]_{\hat{\omega}(\rho)} \right),
\end{equation}
and so matches the
prediction~(\ref{eq:dilaton-shift-conj}).

We will discuss the effect of adding discrete torsion to orbifolds
of this form in section~\ref{sect:ex:trivcentral:dt}.

Next, we will consider several examples of orbifolds in which the
trivially-acting subgroup $K$ is abelian, but not necessarily
central in $\Gamma$.

\subsubsection{Noncentral abelian example:
$[X/{\mathbb H}]$, $K = {\mathbb Z}_4$}

Beginning in this section we will look at examples of orbifolds in which
the trivially-acting subgroup $K$ is not in the center (though is still
abelian).  (Technically, these are ``nonbanded'' abelian gerbes.)
In these cases, the different universes need not have
the same form as one another, and the associated orbits of $\Gamma/K = G$
may contain several representations.

Consider the eight-element group of unit quaternions ${\mathbb H}$,
with trivially-acting $K = \langle i \rangle \cong {\mathbb Z}_4 \subset
{\mathbb H}$.  Decomposition predicts \cite[section 5.4]{Hellerman:2006zs}
\begin{equation}
{\rm CFT}\left( [X/{\mathbb H}] \right) \: = \:
{\rm CFT}\left( X \, \coprod \, [X/{\mathbb Z}_2] \, \coprod \,
[X/{\mathbb Z}_2] \right),
\end{equation}
using the fact that ${\mathbb Z}_2 = {\mathbb H}/\langle i \rangle$ acts
on $K = \langle i \rangle$ by conjugation, which leaves two irreps of
$K$ invariant, and exchanges two others.

From \cite[section 5.4]{Hellerman:2006zs},
\begin{eqnarray}
Z_g\left( [X/{\mathbb H}] \right) & = &
\frac{4^{2g-1}}{|{\mathbb H}|} \left( 2 (\mbox{untwisted sector})
\: + \: 2 (\mbox{all sectors}) \right),
\\
& = &
\frac{4^{2g-1}}{(2)(4)} \left( 2 Z_g(X) \: + \:
2 | {\mathbb Z}_2| Z_g\left( [X/{\mathbb Z}_2] \right)
\right),
\\
& = &
4^{2g-2} Z_g(X) \: + \:
4^{2g-2} (2) Z_g\left( [X/{\mathbb Z}_2] \right),
\\
& = &
4^{2g-2} \left(
 Z_g(X) \: + \:
2 Z_g\left( [X/{\mathbb Z}_2] \right)
\right).
\end{eqnarray}

In this case, since $K = {\mathbb Z}_4$ is abelian,
all its irreducible representations are one-dimensional.
The two  $[X/{\mathbb Z}_2]$ universes are associated
to one-dimensional representations of $K$  which are invariant under the
action of ${\mathbb H}/K = {\mathbb Z}_2$, whereas the
universe $X$ is associated to a pair of one-dimensional representations
that are interchanged by ${\mathbb H}/K = {\mathbb Z}_2$.
In all three cases, representatives of the orbit of $G = {\mathbb Z}_2$
on the space of irreducible representations are one-dimensional,
so in terms of our conjecture~(\ref{eq:dilaton-shift-conj}),
$\dim R_U = 1$ for every universe $U$.
In particular, the result above for the partition function matches
our prediction~(\ref{eq:dilaton-shift-conj}).

\subsubsection{Noncentral abelian example:
$[X/A_4]$, $K = {\mathbb Z}_2 \times {\mathbb Z}_2$}

Consider the group of alternating permutations $A_4$ on four elements.
This has a normal subgroup $K = {\mathbb Z}_2 \times {\mathbb Z}_2$,
which we take to act trivially, and $A_4 / {\mathbb Z}_2 \times
{\mathbb Z}_2 = {\mathbb Z}_3$.

Decomposition predicts \cite[section 5.5]{Hellerman:2006zs}
\begin{equation}
{\rm CFT}\left( [X/A_4] \right) \: = \:
{\rm CFT}\left( [X/{\mathbb Z}_3] \, \coprod \, X \right),
\end{equation}
using the fact that ${\mathbb Z}_3 = A_4 / {\mathbb Z}_2 \times
{\mathbb Z}_2$ leaves invariant the trivial representation of
$K = {\mathbb Z}_2 \times {\mathbb Z}_2$ and permutes the other three.

Analyzing this example at genus $g$ in the same fashion as
\cite[section 5.4]{Hellerman:2006zs},
we find the untwisted sector of the ${\mathbb Z}_3$ orbifold
appears with multiplicity $| {\mathbb Z}_2 \times {\mathbb Z}_2 |^{2g}$
(from multiplying any edge by any element of ${\mathbb Z}_2 \times
{\mathbb Z}_2$),
whereas all twisted sectors appear with multiplicity
$| {\mathbb Z}_2 \times {\mathbb Z}_2 |^{2g-1}$.

Thus, we find
\begin{eqnarray}
Z_g\left([X/A_4]\right) & = &
\frac{4^{2g-1}}{|A_4|} \left(
| {\mathbb Z}_3 | Z_g\left( [X/{\mathbb Z}_3] \right)
\: + \:
(4-1) Z_g(X) \right),
\\
& = &
4^{2g-2} \left( Z_g\left( [X/{\mathbb Z}_3] \right) \: + \:
 Z_g(X) \right).
\end{eqnarray}
In the special case $g=1$, this reduces to
\begin{equation}
Z_1\left( [X/A_4] \right) \: = \: Z_1\left( [X/{\mathbb Z}_3] \right)
\: + \: Z_1(X),
\end{equation}
as expected.

The universe $[X/{\mathbb Z}_3]$ is associated
to a one-dimensional representation of $K$ (which is invariant
under the action of $A_4/K = {\mathbb Z}_3$),
and the universe $X$ is associated to an orbit of $A_4/K$ consisting
of  
three one-dimensional representations of $K$.  In both cases,
representatives of the orbit are one-dimensional, so for both
universes, $\dim R_U = 1$.
In particular, since $|K| = 4$, we see that the result above
matches the prediction~(\ref{eq:dilaton-shift-conj}).

\subsubsection{Noncentral abelian example: $[X/D_n]$, $K = {\mathbb Z}_n$}

Consider the $2n$-element dihedral group $D_n$.  Let its normal
subgroup $K = {\mathbb Z}_n$ act trivially, and recall
$D_n/{\mathbb Z}_n = {\mathbb Z}_2$.

In this case, decomposition predicts \cite[section 5.6]{Hellerman:2006zs}
\begin{equation}
{\rm CFT}\left( [X/D_n] \right) \: = \:
{\rm CFT}\left( \coprod_{\alpha} [X/{\mathbb Z}_2] \, \coprod_{(n-\alpha)/2}
X \right),
\end{equation}
where
\begin{equation}
\alpha \: = \: \left\{ \begin{array}{cl}
1 & n \mbox{ odd}, \\
2 & n \mbox{ even},
\end{array}
\right.
\end{equation}
using the fact that 
${\mathbb Z}_2 = D_n/{\mathbb Z}_n$ leaves $\alpha$ irreps of
$K = {\mathbb Z}_n$ invariant and exchanges the rest in pairs.

Proceeding in the same fashion as \cite[section 5.4]{Hellerman:2006zs},
we compute the partition function at genus $g$.  Here, the untwisted
sector appears with multiplicity $n^{2g}$ (arising from multiplying any
edge by an element of ${\mathbb Z}_{n}$, and all twisted sectors appear
with multiplicity $\alpha n^{2g-1}$.  The partition function then takes
the form
\begin{eqnarray}
Z_g\left( [X/D_n] \right) & = &
\frac{n^{2g-1}}{|D_n|}\left( 
\alpha | {\mathbb Z}_2 | Z_g\left( [X/{\mathbb Z}_2] \right)
\: + \:
(n-\alpha) Z_g(X) \right),
\\
& = &
n^{2g-2} \left[ \alpha Z_g\left( [X/{\mathbb Z}_2] \right)
\: + \: \frac{ (n - \alpha) }{2}  Z_g(X) \right].
\end{eqnarray}
In the special case $g=1$, this reduces to
\begin{equation}
Z_1\left( [X/D_n] \right) \: = \:
\alpha Z_1\left( [X/{\mathbb Z}_2] \right) \: + \:
\frac{n - \alpha}{2} Z_1(X),
\end{equation}
matching the prediction of decomposition.

The universes $[X/{\mathbb Z}_2]$ are each associated 
to one-dimensional representations of 
$K$ (invariant under the action of $D_n/K = {\mathbb Z}_2$),
and the universes $X$ are each associated to orbits of
$D_n/Z_k = {\mathbb Z}_2$ exchanging
pairs of one-dimensional representations of $K$.
As $|K| = n$, we see that the partition functions above match the
predictions~(\ref{eq:dilaton-shift-conj}).

\subsubsection{Nonabelian example: $[X/D_6]$, $K = D_3$}

Next, consider the orbifold $[X/D_6]$ by the twelve-element
dihedral group $D_6$, with trivially-acting normal subgroup $K = 
D_3 \subset D_6$, and recall $D_6 / D_3 = {\mathbb Z}_2$.
In other examples considered so far, $K$ was abelian; here,
since $K$ is nonabelian, it has representations of dimension greater
than one, so the $\dim R_U$ factors in~(\ref{eq:dilaton-shift-conj}) 
will be nontrivial
in the dilaton shifts in higher-genus
partition functions here.

Explicitly, $D_6$ can be presented as generated by $a$, $b$ such that
\begin{equation}
a^2 \: = \: 1 \: = \: b^6,
\: \: \:
a b a \: = \: b^5.
\end{equation}
It is straightforward to check that $z \equiv b^3$ generates the
${\mathbb Z}_2$ center of $D_6$, and the normal subgroup $D_3$
is generated by $\{a, b^2 \}$.

Let $\xi$ denote the nontrivial coset in $D_6/D_3 = {\mathbb Z}_2$.

It will be useful to note that $D_3$ has three irreducible representations,
of dimensions $1$, $1$, and $2$.
In this case, decomposition \cite{Hellerman:2006zs} predicts
\begin{equation} \label{eq:decomp:d6}
{\rm QFT}\left( [X/D_6] \right) \: = \: {\rm QFT} \left( \coprod_3 
[X/{\mathbb Z}_2] \right).
\end{equation}

Specifically, genus $g$ partition functions are predicted 
by~(\ref{eq:dilaton-shift-conj}) to have the form
\begin{equation}
Z_g\left( [X/D_6] \right) \: = \: |K|^{2g-2} \left(
1 + 1 + (2)^{2-2g} \right) Z_g\left( [X/{\mathbb Z}_2 \right),
\end{equation}
since the three representations of $D_3$ have dimension $1$, $1$, $2$.

Now, let us verify this in the partition function for several genera.
The genus $g$ $D_6$ orbifold partition function is already a 
highly intricate combinatorics computation; the fact that it matches
the expression above in examples of genera $g > 1$ will provide a 
strong test of the conjecture~(\ref{eq:dilaton-shift-conj}) (as well as
of decomposition itself).

To that end, it will be useful to characterize commutators
$[a,b] = a b a^{-1} b^{-1}$.

\begin{table}
\begin{center}
\begin{tabular}{c|c|c|c}
Projects to & $[a,b]$ & Pairs & Count \\ \hline
$(1,1)$ & $1$ & $(b^{\rm even}, b^{\rm even})$, $(ab^{2n}, ab^{2n})$,
$(ab^{2n},1)$, $(1,ab^{2n})$ & $18$ \\
$(1,1)$ & $b^2$ & $(ab^{2n}, b^2$), $(b^4, ab^{2n})$,
$(ab^{2i}, ab^{2j})$ ($2i-2j \equiv 2 \mod 6$) & $9$ \\
$(1,1)$ & $b^4$ & $(b^2, ab^{\rm even})$, $(a,b^4)$, $(a,ab^2)$,
$(ab^2,ab^4)$,
 & $9$ \\
& & $(ab^4,a)$, $(ab^2,b^4)$, $(ab^4,b^4)$ & \\
\\
$(1,\xi)$ & $1$ & $(b^{\rm even}, b^{\rm odd})$, $(1,ab^{\rm odd})$,
$(ab^{2n}, b^3)$, $(ab^{2n}, ab^{2n}z)$ & $18$ \\
$(1,\xi)$ & $b^2$ & $(b^4,ab^{\rm odd})$, $(a,b^5)$, $(a,ab)$, $(ab^4,ab^5)$,
& $9$ \\
& & $(ab^2,ab^3)$, $(ab^2, b^5$, $(ab^4,b^5)$ & \\
$(1,\xi)$ & $b^4$ & $(b^2,ab^{\rm odd})$, $(a,b)$, $(a,ab^5)$, $(ab^2,ab)$,
& $9$ \\
& & $(ab^4,ab^3)$, $(ab^2,b)$, $(ab^4,b)$ & \\
\\
$(\xi,\xi)$ & $1$ & $(b^{\rm odd}, b^{\rm odd})$, $(ab^{2k+1}, ab^{2k+1})$,
$(ab^{2n}z,z)$, $(z, ab^{2n}z)$ & $18$ \\
$(\xi,\xi)$ & $b^2$ & $(b,ab^{\rm odd})$, $(ab,ab^5)$, $(ab^3,b^5)$,
$(ab^3,ab)$, & $9$ \\
& & $(ab^5,ab^3)$, $(ab^5,b^5)$, $(ab,b^5)$ & \\
$(\xi,\xi)$ & $b^4$ & $(b^5,ab^{\rm odd})$, $(ab^5,ab)$, $(ab^3,b)$,
$(ab^3,ab^5)$, & $9$ \\
& & $(ab,ab^3)$, $(ab^5,b)$, $(ab,b)$ &
\end{tabular}
\caption{Characterization of ordered pairs of elements of $D_6$.
The first column lists to which pair in $D_6/D_3 = {\mathbb Z}_2$ they project.
The second column is the value of their commutators.
The last column counts the number of entries in the row.
We have omitted ordered pairs corresponding to $(\xi,1)$, as they can
be obtained straightforwardly from the other entries.
\label{table:d6-pairs}
}
\end{center}
\end{table}

Using table~\ref{table:d6-pairs},
the genus-one partition function is given by
\begin{eqnarray}
Z_1\left( [X/D_6] \right) & = &
\frac{1}{|D_6|} \sum_{gh=hg} Z_{g,h},
\\
& = &
\frac{18}{12} \left( 
Z_{1,1} + Z_{1,\xi} + Z_{\xi,1} + Z_{\xi,\xi} \right),
\\
& = &
(3) Z_1\left( [X/{\mathbb Z}_2] \right),
\end{eqnarray}
consistent with the prediction~(\ref{eq:decomp:d6}).

Now, to see dilaton shifts, we have to compute a partition function at genus
different from one, and we will work through the combinatorics here
for $g=2$.
Every sector is determined by group elements $a_{1,2}$, $b_{1,2}$ such
that
\begin{equation}
[a_1, b_1] \, [a_2, b_2] \: = \: 1.
\end{equation}
Let $(a_1 | b_1 | a_2 | b_2 )$ denote group elements,
then sectors that contribute to the $(1 | 1 | 1 | 1)$ sector
of $[X/{\mathbb Z}_2]$ at genus two are, schematically,
\begin{eqnarray}
\lefteqn{
\left( [a_1,b_1] = 1 \right) \left( [a_2,b_2] = 1 \right)
} \\
& & \: + \:
\left( [a_1,b_1) = b^2 \right) \left( [a_2,b_2] = b^4 \right)
\\
& & \: + \:
\left( [a_1,b_1] = b^4 \right)\left( [a_2,b_20 = b^2 \right),
\end{eqnarray}
of which there are
\begin{equation}
(18)(18) \: + \: (9)(9) \: + \: (9)(9) \: = \: 486.
\end{equation}
The counting of other contributions is very similar, and so we find
for the genus two partition function
\begin{eqnarray}
Z_2\left( [X/D_6] \right) & = &
\frac{1}{|D_6|} \sum_{a_i,b_i} Z(a_1,b_1,a_2,b_2),
\\
& = & \frac{486}{12} | {\mathbb Z}_2| Z_2\left( [X/{\mathbb Z}_2] \right),
\\
& = & (81) Z_2\left( [X/{\mathbb Z}_2] \right).
\end{eqnarray}

Now, let us compare to the prediction of~(\ref{eq:dilaton-shift-conj}).
Here, the three universes are associated to representations of dimensions
$1$, $1$, and $2$, so~(\ref{eq:dilaton-shift-conj}) predicts that at genus $g$,
\begin{equation}
Z_g\left( [X/D_6] \right) \: = \: | D_3 |^{2g-2} \left( 
1 + 1 + (2)^{2-2g} \right) Z_g\left( [X/{\mathbb Z}_2] \right).
\end{equation}
At genus $g=2$,
\begin{eqnarray}
Z_2\left( [X/D_6] \right) & = &
| D_3 |^{2} \left( 1 + 1 + (2)^{-2} \right) Z_2\left( [X/{\mathbb Z}_2] \right),
\\
& = & 
(36)\left( 2 + \frac{1}{4} \right) Z_2\left( [X/{\mathbb Z}_2] \right),
\\
& = &
(36)\left( \frac{9}{4} \right) Z_2\left( [X/{\mathbb Z}_2] \right)
\: = \: (81)  Z_2\left( [X/{\mathbb Z}_2] \right),
\end{eqnarray}
matching the result for the genus two partition function above.

Next, let us outline the same consistency test for genus $g=3$.
Here, the partition function is defined by 6-tuples $a_{1,2,3}$, $b_{1,2,3}$,
such that
\begin{equation}
[a_1, b_1] \, [a_2, b_2] \, [a_3, b_3] \: = \: 1.
\end{equation}
The contributions to the $D_6$ orbifold that contribute to any
sector of the ${\mathbb Z}_2$ orbifold are of the form
\begin{eqnarray}
\lefteqn{
\left( [a_1, b_1] = 1 \right) \left( [a_2, b_2] = 1 \right)
\left( [a_3, b_3] = 1 \right)
} \\
& & \: + \:
\left( [a_1,b_1] = b^2 \right) \left( [a_2,b_2] = b^2 \right)
\left( [a_3,b_3] = b^2 \right)
\\
& & \: + \:
\left( [a_1,b_1] = b^4 \right) \left( [a_2,b_2] = b^4 \right)
\left( [a_3,b_3] = b^4 \right)
\\
& & \: + \:
\left( [a_1,b_1] = 1 \right) \left( [a_2,b_2] = b^2 \right)
\left( [a_3,b_3] = b^4 \right)
\\
& & \: + \:
\left( [a_1,b_1] = 1 \right) \left( [a_2,b_2] = b^4 \right)
\left( [a_3,b_3] = b^2 \right)
\\
& & \: + \:
\left( [a_1,b_1] = b^2 \right) \left( [a_2,b_2] = 1 \right)
\left( [a_3,b_3] = b^4 \right)
\\
& & \: + \:
\left( [a_1,b_1] = b^2 \right) \left( [a_2,b_2] = b^4 \right)
\left( [a_3,b_3] = 1 \right)
\\
& & \: + \:
\left( [a_1,b_1] = b^4 \right) \left( [a_2,b_2] = 1 \right)
\left( [a_3,b_3] = b^2 \right)
\\
& & \: + \:
\left( [a_1,b_1] = b^4 \right) \left( [a_2,b_2] = b^2 \right)
\left( [a_3,b_3] = 1 \right),
\end{eqnarray}
and from table~\ref{table:d6-pairs}, there are
\begin{equation}
(18)^3 \: + \: (9)^3 \: + \: (9)^3 \: + \: (18)(9)(9)(6)
\: = \: 16038
\end{equation}
such 6-tuples.  We find for the genus three partition function
\begin{eqnarray}
Z_3\left( [X/D_6] \right) & = &
\frac{1}{|D_6|} \sum_{a_i, b_i} Z(a_1, b_1, a_2, b_2, a_3, b_3),
\\
& = &
\frac{16038}{12} | {\mathbb Z}_2| Z_3\left( [X/{\mathbb Z}_2] \right)
\: = \: (2673) Z_3\left( [X/{\mathbb Z}_2] \right).
\end{eqnarray}

Now, let us compare to the prediction of~(\ref{eq:dilaton-shift-conj}).
Here, the three universes are associated to representations of dimensions
$1$, $1$, and $2$, 
so (\ref{eq:dilaton-shift-conj}) predicts that at genus three,
\begin{eqnarray}
Z_3\left( [X/D_6] \right) & = &
|D_3|^{2(3)-2} \left( 1 + 1 + (2)^{2-2(3)} \right) Z_3\left( [X/{\mathbb Z}_2]
\right),
\\
& = &
(6)^4 \left( 2 + \frac{1}{16} \right)  Z_3\left( [X/{\mathbb Z}_2]
\right),
\\
& = &
(1296)\left( \frac{33}{16} \right)  Z_3\left( [X/{\mathbb Z}_2]
\right)
\: = \: (81)(33)  Z_3\left( [X/{\mathbb Z}_2]
\right),
\\
& = & (2673) Z_3\left( [X/{\mathbb Z}_2]
\right),
\end{eqnarray}
matching the result for the genus three partition function above.
This provides another, rather intricate, test of the
dilaton shift conjecture~(\ref{eq:dilaton-shift-conj}).

\subsubsection{Two-dimensional Dijkgraaf-Witten theory}
\label{sect:2ddw:1}

Two-dimensional Dijkgraaf-Witten theory is the theory of an orbifold
of a point.
The partition function of two-dimensional Dijkgraaf-Witten theory with
orbifold group $G$ is\footnote{
Up to an overall normalization, as previously discussed.  Our conventions
here are consistent with usage elsewhere in this paper.
}
(see \cite{frobenius1,frschur,mednyh}
and also e.g.~\cite{Snyder:2007ns,mulaseyu,Gardiner:2020vjp,Ramgoolam:2022xfk})
\begin{equation}  \label{eq:2ddw:part:1}
Z_g(G) \: = \: |G|^{2g-2} \sum_R \left( \dim R \right)^{2-2g}.
\end{equation}
If there is no discrete torsion, the sum is over all ordinary
irreducible representations $R$, and the partition function above 
can also be written
\begin{equation}
\frac{1}{|G|} | {\rm Hom}(\pi_1,G) |.
\end{equation}
In the presence of discrete torsion, the form of the
partition function $Z_g$ 
is unchanged, though the sum runs over irreducible projective representations
(determined by the discrete torsion), rather than over ordinary
irreducible representations, see 
e.g.~\cite[section C.1, equ'n (C.28)]{Ramgoolam:2022xfk}.
We do not include discrete torsion for the moment, but will return
to Dijkgraaf-Witten theory with discrete torsion
 later in
section~\ref{sect:2ddw:2}.

Now, let us consider the decomposition of two-dimensional Dijkgraaf-Witten
theory for group $G$.
As all of the orbifold group $G$ acts trivially, it decomposes, into a disjoint
union of points indexed by irreducible representations of $G$.
The partition function sum~(\ref{eq:2ddw:part:1})
precisely matches that of
the dilaton shift conjecture~(\ref{eq:dilaton-shift-conj}),
with universes indexed by irreducible
representations of $G$, and with each $X_U$ a point.

In particular, two-dimensional Dijkgraaf-Witten theory provides
a general test of the dilaton shift conjecture~(\ref{eq:dilaton-shift-conj}),
in cases in which $K = G$ is nonabelian.
(Later in section~\ref{sect:2ddw:2} we will apply it as a test in cases
in which the
restriction of discrete torsion to $K$ is nontrivial.)

In passing, the dilaton shift factors at genus zero, namely
\begin{equation}
\frac{ (\dim R)^2 }{ |G|^2 },
\end{equation}
are proportional to the Plancherel measure on the set of irreducible
representations
\begin{equation}
\frac{ (\dim R)^2 }{|G|},
\end{equation}
a point to which we shall return in section~\ref{sect:ds-vs-prob}.

\subsection{Examples in orbifolds with discrete torsion}
\label{sect:orb:dt}

In this section we will study decomposition in orbifolds with discrete
torsion.  Briefly, we will see in numerous examples that the dilaton shift
factors on universes have the same form as previously discussed.

\subsubsection{Conjecture, split into three cases}
\label{sect:orb:dt:conj}

Decomposition in orbifolds with discrete torsion was discussed
in \cite{Robbins:2021ylj}, and has a somewhat more complex form.
Consider an orbifold $[X/\Gamma]_{\omega}$, where $\omega \in H^2(\Gamma,U(1))$
(with trivial action on the coefficients), and where a subgroup
$K \subset \Gamma$ acts trivially.  Describe $G$ by the extension
\begin{equation}
1 \: \longrightarrow \: K \: \stackrel{\iota}{\longrightarrow} \: \Gamma
\: \stackrel{\pi}{\longrightarrow} \: \Gamma/K \: \longrightarrow \: 1.
\end{equation}
Then, decomposition is described in terms of the three cases, which we
briefly review next.
\begin{enumerate}
\item In the case $\iota^* \omega \neq 0$, then,
\begin{equation}
{\rm QFT}\left( [X/\Gamma]_{\omega} \right) \: = \:
{\rm QFT} \left( \left[ \frac{ X \times \hat{K}_{\iota^* \omega} }{ \Gamma/K }
\right]_{\hat{\omega}} \right),
\end{equation}
where $\hat{K}_{\iota^* \omega}$ denotes the set of irreducible projective
representations of $K$, twisted by $\iota^* \omega$, and $\hat{\omega}$
is discrete torsion on the factors.
\item Suppose that $\iota^* \omega = 0$, then there is an exact sequence
\begin{equation}
H^2( \Gamma/K, U(1)) \: \stackrel{\pi^*}{\longrightarrow} \:  L \: 
\stackrel{\beta}{\longrightarrow} \:
H^1( \Gamma/K, H^1(K,U(1))),
\end{equation}
where
\begin{equation}
L \: = \: {\rm Ker}\, \iota^*: \: H^2(\Gamma,U(1)) \: \longrightarrow \:
H^2(K,U(1)).
\end{equation}
If $\beta(\omega) \neq 0$ and, for simplicity, $K$ is in the center of $\Gamma$,
then
\begin{equation}
{\rm QFT}\left( [X/\Gamma]_{\omega} \right) \: = \:
{\rm QFT}\left( \left[ \frac{ X \times {\rm Coker}\left( \beta(\omega) \right)
}{ {\rm Ker}(\beta(\omega)) } \right]_{\hat{\omega}} \right),
\end{equation}
where we interpret $\beta(\omega)$ as a homomorphism
$\Gamma/K \rightarrow \hat{K}$, and $\hat{\omega}$ denotes discrete torsion on
factors.
\item The final case is that $\iota^* \omega = 0$ and $\beta(\omega) = 0$.
Then, there exists $\overline{\omega} \in H^2(\Gamma/K,U(1))$ such that $\omega
 = \pi^* \overline{\omega}$, and
\begin{equation}
{\rm QFT}\left( [X/\Gamma]_{\omega} \right) \: = \:
{\rm QFT}\left( \left[ \frac{ X \times \hat{K} }{ \Gamma/K } \right]_{\hat{\omega}}
\right).
\end{equation}
In this case, the effect of $\overline{\omega}$ is to shift the discrete
torsion $\hat{\omega}$ on factors relative to the case $\omega=0$.
\end{enumerate}

We conjecture that the dilaton shift factors have the same form in the
case in which the orbifold has discrete torsion, namely that the genus
$g$ partition function is
\begin{equation}
Z_g\left( [X/\Gamma]_{\omega} \right) \: = \: 
\sum_U \left( \frac{\dim R_U}{|K|} \right)^{2-2g} Z_g(X_U),
\end{equation}
where $X_U$ denotes the theory for universe $U$, as determined by
the statements above, and where $Z_g$ is normalized in the same way
discussed in section~\ref{sect:orb:nodt:conj}.  (As discussed in that
section, we could also phrase this more invariantly in terms of
relative dilaton shifts.)
However, the interpretation of $R_U$ differs between the three cases,
as we describe below:
\begin{enumerate}
\item In the case $\iota^* \omega \neq 0$, $R_U$ denotes a
$\iota^*\omega$-twisted projective irreducible representation in
$\hat{K}_{\iota^*\omega}$ in the
orbit of $\Gamma/K$.
\item In the case $\iota^* \omega = 0$ and $\beta(\omega) \neq 0$,
$R_U$ denotes an ordinary irreducible representation in
Coker$(\beta(\omega)) \subset \hat{K}$ in the orbit of
Ker$(\beta(\omega))$, and in the trivial case Coker$(\beta(\omega)) = 0$, 
we take dim $R_U = 1$.
\item In the case $\iota^* \omega = 0$ and $\beta(\omega) = 0$,
$R_U$ denotes an ordinary irreducible representation of $K$,
an element of $\hat{K}$, in the orbit of $\Gamma/K$.
\end{enumerate}
In passing, note that the case of vanishing discrete torsion is
part of case (3), and it is easy to see that the conjecture for that
case correctly specializes to previous results.

We will check the conjecture above in examples of each type given,
in the next several subsections.

\subsubsection{Example of case (1): Two-dimensional Dijkgraaf-Witten theory}
\label{sect:2ddw:2}

Previously in section~\ref{sect:2ddw:1} we discussed two-dimensional
Dijkgraaf-Witten theory (without discrete torsion) as an example in which 
the conjecture for dilaton shift factors could be checked.

Since the entire orbifold group acts trivially, two-dimensional Dijkgraaf-Witten
theory with discrete torsion is an example of case (1).

Our analysis is nearly identical to the case without discrete torsion.
Here, the partition function of Dijkgraaf-Witten theory for finite
group $G$ is well-known to be given by
\begin{equation}
Z_g(G) \: = \: \sum_R \left( \frac{\dim R}{|G|} \right)^{2-2g}
Z_g({\rm point}),
\end{equation}
where the sum is over irreducible projective representations
(twisted by the $\omega \in H^2(G,U(1))$ corresponding to
discrete torsion).
This immediately reproduces the prediction of the
conjecture of section~\ref{sect:orb:dt:conj}.

\subsubsection{Example of case (2): $[X/{\mathbb Z}_2 \times {\mathbb Z}_2]$,
$K = {\mathbb Z}_2$}

Consider the case of the orbifold $[X/{\mathbb Z}_2 \times {\mathbb Z}_2]$
with nontrivial discrete torsion, and with one trivially-acting ${\mathbb Z}_2$
factor.  This was discussed (at genus one) in
\cite[section 5.1]{Robbins:2021ylj}.  Briefly, the analysis there predicted
\begin{equation}
{\rm QFT}\left( [X/{\mathbb Z}_2 \times {\mathbb Z}_2]_{\omega} \right)
\: = \: {\rm QFT}\left(X\right).
\end{equation}
In terms of the conjecture, it was argued in \cite[section 5.1]{Robbins:2021ylj}
that $\iota^* \omega = 0$ and $\beta(\omega)$ is an isomorphism, hence
\begin{equation}
{\rm Ker}\left( \beta(\omega) \right) \: = \: 0 \: = \:
{\rm Coker}\left( \beta(\omega) \right).
\end{equation}

Discrete torsion phase factors on Riemann surfaces of genus greater than one
can be found in e.g.~\cite[equ'n (15)]{Aspinwall:2000xv}, \cite{Bantay:2000eq}.
In the case that the orbifold group $G$ is abelian, those phase factors
reduce to a product of genus-one phase factors, so that on a genus $g$
Riemann surface, the phase factor is
\begin{equation}
\prod_{i=1}^g \frac{ \omega(a_i,b_i) }{ \omega(b_i,a_i) },
\end{equation}
where $\{a_1, b_1, a_2, b_2, \cdots, a_g, b_g \}$ are the elements of
$G$ defining a given genus $g$ twisted sector.

Now, write the two generators of ${\mathbb Z}_2 \times {\mathbb Z}_2$ as
$x$, $y$, where $x$ acts trivially on $X$, and $y$ acts nontrivially on $X$.

Let $Z(a_1, b_1, a_2, b_2, \cdots, b_g)$ denote the twisted sector arising
from a polygon with sides labelled by the group elements $a_1, \cdots, b_g$.
Generalizing the computations of \cite{Robbins:2021ylj}, it is straightforward
to check that contributions from $Z(a_1, b_1, \cdots, b_g)$ with any
$a_i \not\in \{1, x\}$ or $b_i \not\in \{1, x\}$ cancel out.
That leaves $4^g$ remaining twisted sectors, and since $x$ acts trivially,
they can all be identified with $Z_g(X)$.  Including the overall normalization
of $1/| {\mathbb Z}_2 \times {\mathbb Z}_2|$, we find that the genus
$g$ partition function is given by
\begin{equation}
Z_g\left( [X/{\mathbb Z}_2 \times {\mathbb Z}_2]_{\omega} \right)
\: = \: \frac{4^g}{| {\mathbb Z}_2 \times {\mathbb Z}_2|} Z_g(X)
\: = \: 4^{g-1} Z_g(X).
\end{equation}

Now, let us compare to the prediction of section~\ref{sect:orb:dt:conj}.
Here, since Coker$(\beta(\omega)) = 0$, we take dim $R_U = 1$,
so the dilaton shift factor on each universe is
\begin{equation}
\left( \frac{ \dim R_U }{ |K| } \right)^{2-2g} \: = \:
\left( |K|^2 \right)^{g-1} \: = \: 4^{g-1},
\end{equation}
as $K = {\mathbb Z}_2$.  Since both the kernel and cokernel of $\beta$
vanish, there is only one universe, so the prediction is predicted
to be
\begin{equation}
Z_g\left(  [X/ {\mathbb Z}_2 \times {\mathbb Z}_2]_{\omega} \right)
\: = \: 4^{g-1} Z_g(X),
\end{equation}
which matches the result above.

\subsubsection{Example of case (2): $[X/{\mathbb Z}_2 \times {\mathbb Z}_4]$,
$K = {\mathbb Z}_4$}

Consider the case of the orbifold $[X/{\mathbb Z}_2 \times {\mathbb Z}_4]$
with nontrivial discrete torsion, and with trivially-acting ${\mathbb Z}_4$.
This was discussed (at genus one) in
\cite[section 5.3]{Robbins:2021ylj}. 
Briefly, the analysis there predicted
\begin{equation}
{\rm QFT}\left( [X/{\mathbb Z}_2 \times {\mathbb Z}_4]_{\omega} \right)
\: = \:
{\rm QFT}\left( X \coprod X \right).
\end{equation}
In terms of the conjecture, it was argued in
\cite[section 5.3]{Robbins:2021ylj} that 
\begin{equation}
{\rm Ker}\left( \beta(\omega) \right) \: = \: 0, \: \: \:
{\rm Coker}\left( \beta(\omega) \right) \: = \: {\mathbb Z}_2.
\end{equation}

As in the previous subsection, since the orbifold group is abelian,
discrete torsion on a genus $g$ Riemann surface is the product of discrete
torsion on factorized genus one surfaces.  

Write the generator of ${\mathbb Z}_2$ as $a$, and the generator of
${\mathbb Z}_4$ as $b$.  We assume that $b$ acts trivially.

Let $Z(a_1, b_1, \cdots, b_g)$ denote the twisted sector arising from a polygon
with sides labelled by group elements $a_1, \cdots, b_g \in {\mathbb Z}_2 \times
{\mathbb Z}_4$.  Generalizing the computations of 
\cite[section 5.3]{Robbins:2021ylj},
it is straightforward to check that most sectors cancel out
(due to signs arising from discrete torsion phase factors), with the
exception of sectors in which all of the $a_i, b_i$ are powers of $b$.
At genus $g$, there are $2g$ such factors, and as $b$ acts trivially,
the partition function is then
\begin{equation}
Z_g\left( [X/{\mathbb Z}_2 \times {\mathbb Z}_4]_{\omega} \right)
\: = \: \frac{4^{2g}}{| {\mathbb Z}_2 \times {\mathbb Z}_4|}
Z(X) \: = \: 4^{2g-2} Z\left( X \coprod X \right).
\end{equation}

Now, let us compare to the prediction of section~\ref{sect:orb:dt:conj}.
Here, since Coker$(\beta(\omega)) = {\mathbb Z}_2$, 
dim $R_U = 1$, so the dilaton shift factor on each universe is predicted
to be
\begin{equation}
\left( \frac{ \dim R_U }{|K|} \right)^{2-2g} \: = \: \left(
|K| \right)^{2g-2} \: = \: 4^{2g-2},
\end{equation}
as $K = {\mathbb Z}_4$.  Thus, the prediction is
\begin{equation}
Z_g\left( [X/{\mathbb Z}_2 \times {\mathbb Z}_4]_{\omega} \right)
\: = \:
 4^{2g-2} Z\left( X \coprod X \right),
\end{equation}
which matches the result above.

\subsubsection{Example of case (2): $[X/{\mathbb Z}_2 \times {\mathbb Z}_4]$,
$K = {\mathbb Z}_2$}

Consider the case of the orbifold $[X/{\mathbb Z}_2 \times {\mathbb Z}_4]$
with nontrivial discrete torsion, and with trivially-acting ${\mathbb Z}_2$.
This was discussed in \cite[section 5.4]{Robbins:2021ylj}.  
Briefly, the analysis there predicted
\begin{equation}
{\rm QFT}\left( [X/{\mathbb Z}_2 \times {\mathbb Z}_2]_{\omega} \right)
\: = \: {\rm QFT}\left( [X/{\mathbb Z}_2] \right).
\end{equation}
In terms of the conjecture, it was argued in  
\cite[section 5.4]{Robbins:2021ylj}  that
\begin{equation}
{\rm Ker}\left( \beta(\omega) \right) \: = \: {\mathbb Z}_2, \: \: \:
{\rm Coker}\left( \beta(\omega) \right) \: = \: 0.
\end{equation}

As in the previous subsection, since the orbifold group is abelian,
discrete torsion on a genus $g$ Riemann surface is the product of discrete
torsion on factorized genus one surfaces.

Write the generator of ${\mathbb Z}_2$ as $a$, and the generator of
${\mathbb Z}_4$ as $b$.  We assume that $a$ acts trivially.

Let $Z(a_1, b_1, \cdots, b_g)$ denote the twisted sector arising from a polygon
with sides labelled by group elements $a_1, \cdots, b_g \in {\mathbb Z}_2 \times
{\mathbb Z}_4$.  First, consider a genus one computation.
From the discrete torsion phase factors in \cite[table D.2]{Robbins:2021ylj}. 
it is straightforward to check that, for example,
\begin{equation}
\epsilon(a b^i, b^j) \: = \: (-)^j, \: \: \:
\epsilon(b^i, ab^j) \: = \: (-)^i, \: \: \:
\epsilon(ab^i, ab^j) \: = \: (-)^{i+j}.
\end{equation}
Given that $a$ acts trivially, it is then straightforward to compute that
the genus $g$ partition function is given by
\begin{eqnarray}
\lefteqn{
Z_g\left( [X/{\mathbb Z}_2 \times {\mathbb Z}_4]_{\omega} \right)
}  \\
& = & \frac{1}{| {\mathbb Z}_2 \times {\mathbb Z}_4|} 
\sum_{i_1, j_1, i_2, \cdots, j_g = 0}^3 
Z(b^{i_1}, b^{j_1}, b^{i_2}, \cdots, b^{j_g}) \left(
1 + (-)^i + (-)^j + (-)^{i+j} \right)^g.
\nonumber
\end{eqnarray}
Now,
\begin{equation}
1 + (-)^i + (-)^j + (-)^{i+j}  \: = \: \left\{ \begin{array}{cl}
4 & \mbox{$i$ even and $j$ even},
\\
0 & \mbox{else},
\end{array}
\right.
\end{equation}
hence
\begin{eqnarray}
Z_g\left( [X/{\mathbb Z}_2 \times {\mathbb Z}_4]_{\omega} \right)
& = &
\frac{4^g}{| {\mathbb Z}_2 \times {\mathbb Z}_4|} 
| {\mathbb Z}_2 | \, Z_g\left( [X/ \langle b^2 \rangle ] \right),
\\
& = &
4^{g-1} \,  Z_g\left( [X/ {\mathbb Z}_2 ] \right).
\end{eqnarray}

Now, let us compare to the prediction of section~\ref{sect:orb:dt:conj}.
Here, since Coker$(\beta(\omega)) = 0$,
dim $R_U = 1$, so the dilaton shift factor on each universe is predicted
to be
\begin{equation}
\left( \frac{ \dim R_U }{|K|} \right)^{2-2g} \: = \: \left(
|K| \right)^{2g-2} \: = \: 2^{2g-2} \: = \: 4^{g-1},
\end{equation}
as $K = {\mathbb Z}_2$.  Thus, the prediction is
\begin{equation}
Z_g\left( [X/{\mathbb Z}_2 \times {\mathbb Z}_4]_{\omega} \right)
\: = \:
4^{g-1} Z_g\left( [X/ {\mathbb Z}_2 ] \right),
\end{equation}
which matches the result above.

\subsubsection{Examples of case (3): Trivially-acting central subgroups}
\label{sect:ex:trivcentral:dt}

Previously in subsection~\ref{sect:ex:trivcentral} we discussed examples
of dilaton shift factors in decomposition in orbifolds in which the
trivially-acting subgroup $K$ is central in the orbifold group $\Gamma$.  
In this section we extend those remarks
to the case in which the decomposing orbifold has discrete torsion,
in case (3), so that the discrete torsion in the $\Gamma$ orbifold
is a pullback from discrete torsion in a $\Gamma/K$ orbifold.

Let $[\omega] \in H^2(\Gamma,U(1))$ denote discrete torsion in the
$\Gamma$ orbifold, and assume that
$\omega = \pi^* \tilde{\omega}$ for some $[\tilde{\omega}] \in
H^2(G,U(1))$.
As discussed in \cite{Robbins:2021ylj}, in this case, the effect of
$\omega$ is merely to shift the discrete torsion on universes by
$\tilde{\omega}$; the decomposition is otherwise unchanged.
If
we let $\epsilon_g(a_i, b_i)$ denote the genus-$g$ discrete torsion
phases in a given sector defined by $a_i, b_i \in \Gamma$,
then in this case,
\begin{equation}
\epsilon_g(a_i, b_i) \: = \: 
\epsilon_g(\overline{a}_i, \overline{b}_i)
\end{equation}
for $\overline{a}_i, \overline{b}_i \in G$,
and so $\epsilon_g$ can be completely absorbed into
$Z(\overline{a}_i,\overline{b}_i)$.
In this case,
we see immediately that the dilaton shifts are identical to those
without the discrete torsion.  For example, assuming again that
$K$ is central, the genus $g$ partition function is
\begin{equation}
Z_g \: = \: |K|^{2g-2} \sum_U Z'_g(X_U),
\end{equation}
where $Z'_g(X_U)$ is the genus-$g$ partition function of
universe $X_U$, with discrete torsion $\tilde{\omega}$.
This confirms our conjecture of section~\ref{sect:orb:dt:conj} for this
case of discrete torsion in the $\Gamma$ orbifold,
since all representations $R_U$ of $K$ are one-dimensional for $K$
central.

Next, we will discuss some concrete examples of this form.

\subsubsection{Example of case (3): $[X/ {\mathbb Z}_2 \times {\mathbb Z}_4]$,
$K = {\mathbb Z}_2$}

Consider the orbifold $[X/ {\mathbb Z}_2 \times {\mathbb Z}_4]$
with nontrivial discrete torsion, and one trivially-acting ${\mathbb Z}_2$
factor.  Write ${\mathbb Z}_2 = \langle x \rangle$,
${\mathbb Z}_4 = \langle y \rangle$, and take the trivially-acting
$K = \langle y^2 \rangle \cong {\mathbb Z}_2$.
This was discussed (at genus one) in
\cite[section 6.1]{Robbins:2021ylj}.  Briefly, the analysis there
argued that $\iota^* \omega = 0$, $\omega = \pi^* \overline{\omega}$
for $\overline{\omega} \in H^2( {\mathbb Z}_2 \times {\mathbb Z}_2, U(1))$,
and that
\begin{equation}
{\rm QFT}\left( [X/ {\mathbb Z}_2 \times {\mathbb Z}_4]_{\omega} \right)
\: = \: {\rm QFT}\left( \coprod_2 [ X / {\mathbb Z}_2 \times {\mathbb Z}_2]_{
\overline{\omega}} \right).
\end{equation}

As noted previously, in abelian orbifolds, discrete torsion phases
on a genus $g$ Riemann surface are the
product of discrete torsion on factorized genus one surfaces.
The genus one phases can be found in \cite[table D.2]{Robbins:2021ylj},
and from the table there it is easy to see that multiplication by $y^2$
does not change the discrete torsion phase.  It is then straightforward
to compute
\begin{eqnarray}
Z_g\left( [X/ {\mathbb Z}_2 \times {\mathbb Z}_4]_{\omega} \right)
& = &
\frac{2^{2g}}{| {\mathbb Z}_2 \times {\mathbb Z}_4|} \,
| {\mathbb Z}_2 \times {\mathbb Z}_2| \,
Z_g\left( [ X/ {\mathbb Z}_2 \times {\mathbb Z}_2]_{\overline{\omega}}
\right),
\\
& = & \frac{2^{2g}}{2} \, Z_g\left( [ X/ {\mathbb Z}_2 \times {\mathbb Z}_2]_{\overline{\omega}}
\right),
\\
& = & 2^{2g-2} \,  Z_g\left( [ X/ {\mathbb Z}_2 \times {\mathbb Z}_2]_{\overline{\omega}}
\right).
\end{eqnarray}

Now, let us compare to the prediction of section~\ref{sect:orb:dt:conj}.
Here, $|K| = 2$ and $\dim R_U = 1$ for each universe, so the dilaton shift
factor is predicted to be
\begin{equation}
\left( \frac{ \dim R_U }{|K|} \right)^{2-2g} \: = \:
\left( |K| \right)^{2g-2} \: = \: 2^{2g-2},
\end{equation}
as $K = {\mathbb Z}_2$.  Thus, the prediction is
\begin{equation}
Z_g\left( [X/ {\mathbb Z}_2 \times {\mathbb Z}_4]_{\omega} \right)
\: = \:  2^{2g-2} \,  Z_g\left( [ X/ {\mathbb Z}_2 \times {\mathbb Z}_2]_{\overline{\omega}}
\right),
\end{equation}
which matches the result above.

\subsubsection{Example of case (3): $[X/{\mathbb Z}_4 \rtimes {\mathbb Z}_4]$,
$K = {\mathbb Z}_2$ }

Consider next the case of the orbifold 
$[X/{\mathbb Z}_4 \rtimes {\mathbb Z}_4]$ (the semidirect product
of two copies of ${\mathbb Z}_4$), with discrete torsion,
and a trivially-acting ${\mathbb Z}_2$ in the center of
${\mathbb Z}_4 \rtimes {\mathbb Z}_4$.
This was discussed (at genus one) in
\cite[section 6.2]{Robbins:2021ylj}.
We will use the same notation as in \cite[appendix D.4]{Robbins:2021ylj};
for example,
we let $x$, $y$ denote the gnerators of the two copies of ${\mathbb Z}_4$,
and $K = \langle y^2 \rangle$, so that ${\mathbb Z}_4 \rtimes {\mathbb Z}_4
/ K = D_4$.  
Briefly, the analysis of \cite[section 6.2]{Robbins:2021ylj} predicted that
\begin{equation}
{\rm QFT}\left( [X/{\mathbb Z}_4 \rtimes {\mathbb Z}_4]_{\omega} \right)
\: = \:
{\rm QFT}\left( \coprod_2 [X/D_4]_{\overline{\omega}} \right),
\end{equation}
where $\overline{\omega} \in H^2(D_4,U(1))$ and 
$\omega = \pi^* \overline{\omega}$.

Following the analysis of \cite[section 6.2]{Robbins:2021ylj},
and using the fact that $K$ lies in the center, it is straightforward
to compute that 
\begin{eqnarray}
Z_g\left( [X/{\mathbb Z}_4 \rtimes {\mathbb Z}_4]_{\omega} \right)
& = &
\frac{2^{2g}}{ | {\mathbb Z}_4 \rtimes {\mathbb Z}_4 | }
| D_4 | \, Z_g\left( [X / D_4]_{\overline{\omega}} \right),
\\
& = &
\frac{2^{2g}}{2}  \, Z_g\left( [X / D_4]_{\overline{\omega}} \right),
\\
& = &
2^{2g-2}  \, Z_g\left( \coprod_2 [X / D_4]_{\overline{\omega}} \right).
\end{eqnarray}

Now, let us compare to the prediction of section~\ref{sect:orb:dt:conj}.
Here, $|K| = 2$ and $\dim R_U = 1$ for each universe, so the dilaton shift
factor is predicted to be
\begin{equation}
\left( \frac{ \dim R_U }{|K|} \right)^{2-2g} \: = \:
\left( |K| \right)^{2g-2} \: = \: 2^{2g-2},
\end{equation}
as $K = {\mathbb Z}_2$.  Thus, the prediction is
\begin{equation}
Z_g\left( [X/{\mathbb Z}_4 \rtimes {\mathbb Z}_4]_{\omega} \right)
\: = \:
2^{2g-2}  \, Z_g\left(\coprod_2  [X / D_4]_{\overline{\omega}} \right),
\end{equation}
which matches the result above.

\subsection{Examples with quantum symmetries}
\label{sect:exs:qs}

Next, let us consider examples with quantum symmetries
in the sense of \cite{Robbins:2021ibx}.
Consider orbifolds $[X/\Gamma]$ where a central subgroup
$K \subset \Gamma$ acts trivially, and $G = \Gamma/K$.
The quantum symmetry, as the term is used in \cite{Robbins:2021ibx},
is an element $B \in H^1(G, H^1(K,U(1))$ (generalizing older notions
of quantum symmetries), that provides a relative phase between sectors.
It was argued in \cite{Robbins:2021ibx} that in the presence of such $B$,
\begin{equation}
{\rm QFT}\left( [X/\Gamma]_B \right) \: = \: {\rm QFT}\left(
\left[ \frac{ X \times \widehat{ {\rm Coker}\, B} }{ {\rm Ker}\, B } 
\right] \right),
\end{equation}

This notion of quantum symmetries specializes to both ordinary
quantum symmetries and to results on discrete torsion,
as discussed in \cite{Robbins:2021ibx}..
For example, as discussed in \cite[section 5.1]{Robbins:2021ylj},
the ordinary quantum symmetry of a ${\mathbb Z}_2$ orbifold, and the
fact that orbifolding by the quantum symmetry returns the original
theory, can be understood as decomposition in a
${\mathbb Z}_2 \times {\mathbb Z}_2$ orbifold with discrete torsion.
In terms of orbifolds with discrete torsion, the pertinent decomposition
is of case (2) in the classification of section~\ref{sect:orb:dt}.
The decomposition statement above therefore generalizes those results.

We conjecture that the dilaton shift factors have the same form as we have
seen in the previous examples, namely
\begin{equation}  \label{eq:dilaton-shift-conj-qs}
Z_g\left( [X/\Gamma]_{\omega} \right) \: = \: 
\sum_U \left( \frac{\dim R_U}{|K|} \right)^{2-2g} Z_g(X_U),
\end{equation}
where $X_U$ denotes the theory for universe $U$,
and $R_U$ is an irreducible representation of Coker $B$, in the orbit of
Ker $B$.  (More invariantly, these dilaton shifts should be understood
as relative dilaton shifts between the partition functions for universe
$U$ and for the ambient theory, as discussed in 
section~\ref{sect:orb:nodt:conj}.)

We will work through several examples.

For use in computing examples, we make an observation next.
Describe the contributions to a genus $g$ partition function of $[X/\Gamma]$
as
\begin{equation}
(a_1 | b_1 | a_2 | b_2 | \cdots | a_g | b_g )
\end{equation}
for $a_{1-g}, b_{1-g} \in \Gamma$, where
\begin{equation}
\prod_{i=1}^g [a_i, b_i] \: = \: 1,
\end{equation}
then a quantum symmetry $B \in H^1(G, H^1(K,U(1)))$ relates sectors as
\begin{eqnarray}
(a_1 z | b_1 | a_2 | b_2 | \cdots | b_g) & = &
B(\pi(b_1), z) \, (a_1 | b_1 | a_2 | b_2 | \cdots | a_g | b_g ),
\\
(a_1 | b_1 z | a_2 | b_2 | \cdots |  b_g) & = &
B(\pi(a_1), z)^{-1} \, (a_1 | b_1 | a_2 | b_2 | \cdots | a_g | b_g ),
\\
(a_1 | b_1 | a_2 z | b_2 | \cdots | b_g) & = &
B(\pi(b_2), z) \, (a_1 | b_1 | a_2 | b_2 | \cdots | a_g | b_g ),
\\
(a_1 | b_1 | a_2 | b_2 z | \cdots | b_g ) & = &
B(\pi(a_2), z)^{-1} \, (a_1 | b_1 | a_2 | b_2 | \cdots | a_g | b_g ),
\end{eqnarray}
and so forth, where $z \in K$.

Now, consider the case of a ${\mathbb Z}_4$ orbifold with
trivially-acting $K = {\mathbb Z}_2 \subset {\mathbb Z}_4$,
and a nontrivial quantum symmetry in Hom$({\mathbb Z}_2,
\hat{\mathbb Z}_2)$.  This was discussed in
\cite[section 4.1.1]{Robbins:2021ibx}, which argued that
\begin{equation}
{\rm QFT}\left( [X/{\mathbb Z}_4] \right) \: = \:
{\rm QFT}\left( X \right).
\end{equation}

From the general prediction~(\ref{eq:dilaton-shift-conj-qs}),
since
there is only one universe, which is associated to a representation of
dimension $1$, it should be the case that
\begin{equation}
Z_g\left( [X/{\mathbb Z}_4] \right) \: = \: 
|K|^{2g-2} Z(X)
\: = \:
2^{2g-2} Z(X).
\end{equation}

Label the elements of ${\mathbb Z}_4$ by $i \in \{0, \cdots, 3\}$
as in \cite[section 4.1.1]{Robbins:2021ibx}.
Then, for example,
\begin{equation}
(a_1 + 2 | b_1 | a_2 | b_2 | \cdots | b_g ) \: = \:
(-)^{b_1} (a_1 | b_1 | a_2 | b_2 | \cdots | b_g ),
\end{equation}
It is straightforward to check
that if any of $a_{1-g}$, $b_{1-g} \in \{0, \cdots, 3\}$ are odd,
then contributions from that sectors cancel out.
For example, if $a_1$ is odd, the contributions from the sector
with $b_1+2$ cancels out the contribution from that sector.
As a result, the only sectors that contribute to the
partition function have all of $a_{1-g}$, $b_{1-g}$ even, and these
all match the contribution from the $(0|0|0|\cdots|0)$ sector,
which is exactly $Z(X)$.

Counting contributions, we have
\begin{equation}
Z_g\left( [X/{\mathbb Z}_4]_B \right) \: = \:
\frac{1}{| {\mathbb Z}_4 | } 2^{2g} Z(X) \: = \:
2^{2g-2} Z(X),
\end{equation}
which confirms the prediction~(\ref{eq:dilaton-shift-conj-qs}).

Next, let us consider the orbifold $[X/{\mathbb Z}_{2k}]_B$,
for $k$ even,
as discussed in \cite[section 4.1.2]{Robbins:2021ibx}.
Here, $K = {\mathbb Z}_k \subset {\mathbb Z}_{2k}$ acts trivially,
with a nontrivial quantum symmetry $B \in H^1(G,H^1((K,U(1))))$ such that,
if $x$ denotes the generator of ${\mathbb Z}_{2k}$, so that
$x^2$ generates $K = {\mathbb Z}_k$, then
\begin{equation}
{\scriptstyle x^2} \square_x \: = \: - \left(
{\scriptstyle 1} \square_x \right).
\end{equation}

For this theory, decomposition predicts
\begin{equation}
{\rm QFT}\left( [X/{\mathbb Z}_{2k}]_B \right) \: = \:
{\rm QFT}\left( \coprod_{k/2} X \right).
\end{equation}
Furthermore, the dilaton shift conjecture~(\ref{eq:dilaton-shift-conj-qs})
predicts
\begin{equation}
Z_g\left( [X/{\mathbb Z}_{2k}]_B \right) \: = \:
|K|^{2g-2} Z_g\left( \coprod_{k/2} X \right),
\end{equation}
since each universe is associated with a dimension-one irreducible
representation of $K$.

It is straightforward to check this prediction explicitly, as the details
are not dissimilar to the previous example.  Much as happened there,
if we enumerate group elements by integers in $\{0, \cdots, 2k-1\}$,
then twisted sectors
\begin{equation}
( a_1 | b_1 | a_2 | \cdots | b_g )
\end{equation}
cancel out whenever any of the $a_i$ or $b_i$ are odd.  The only surviving
sectors are those for which all $a_i$ and $b_i$ are even -- in which
case the boundary conditions are trivial, so that the sector is equivalent
to the partition function of a sigma model on $X$.  Since there
are a total of $k^{2g}$ such sectors in which all the $a_i$ and $b_i$
are even, this means that
\begin{eqnarray}
Z_g\left( [X/{\mathbb Z}_{2k}]_B \right)
& = &
\frac{1}{| {\mathbb Z}_{2k} |} \sum_{a_i, b_i}
(a_1 | b_1 | a_2 | \cdots | b_g ),
\\
& = &
\frac{1}{| {\mathbb Z}_{2k} |}  k^{2g}
Z(X),
\\
& = &
k^{2g-2} \frac{k}{2} Z(X) \: = \: k^{2g-2} Z\left( \coprod_{k/2} X \right),
\end{eqnarray}
again matching the prediction of~(\ref{eq:dilaton-shift-conj-qs}).

\subsection{Examples in gauge theories}
\label{sect:exs:gauge}

\subsubsection{Pure Yang-Mills: decomposition along center symmetries}

In this section we will look at the decomposition of two-dimensional
pure Yang-Mills theory along a center one-form symmetry,
as described in detail in \cite[section 2.4]{Sharpe:2014tca}.
Two-dimensional pure Yang-Mills theories have a second decomposition,
to invertible field theories \cite{Nguyen:2021yld,Nguyen:2021naa},
which we will discuss in this context
in section~\ref{sect:ym:inv}.  We will recover the same dilaton shifts as
before, and also find a connection between those dilaton shifts and
moduli space volumes.

Consider pure Yang-Mills theory in two dimensions with gauge
group $G$, which we will assume to be semisimple,
and which has (finite) center $K$.
As discussed in \cite[section 2.4]{Sharpe:2014tca},
this decomposes into a sum of $G/K$ gauge
theories with variable discrete theta angles $\lambda \in \hat{K}$,
schematically as
\begin{equation}
{\rm QFT}(G) \: = \:  \oplus_{\lambda \in \hat{K}}
{\rm QFT}(G/K, \lambda).
\end{equation}

Now, let us compare partition functions.
Up to overall factors, the
partition function of
pure $G$ Yang-Mills on a Riemann surface $\Sigma$ of genus $g$ is
\cite{Migdal:1975zg,Drouffe:1978py,Lang:1981rj,Menotti:1981ry,Rusakov:1990rs,Witten:1991we,Witten:1992xu,Thompson:1992hv,Blau:1991mp,Blau:1993hj,Cordes:1994fc}
\begin{equation} 
Z(G) \: \propto \: \sum_R (\dim R)^{2-2g} \exp\left( - A C_2(R) \right).
\end{equation}
For our purposes, we need to fix a normalization, which was determined in
\cite{Witten:1991we},
\cite[equ'n (4.18)]{Witten:1992xu} through a
careful analysis of the zero-area limit and Reidemeister-Ray-Singer torsion
to be
\begin{equation}  \label{eq:2dym:part:exact}
Z(G) \: = \: \left( \frac{ {\rm Vol}(G) }{ (2\pi)^{{\rm dim}\, G} } 
\right)^{2g-2} \sum_R \left( {\rm dim}\, R \right)^{2-2g}
\exp\left( - A C_2(R) \right).
\end{equation}

Now, define $Z(G/K,\lambda)$ to be the partition function of the
corresponding $G/K$ Yang-Mills theory with discrete theta angle
$\lambda$ (for which the sum is restricted to representations of
corresponding $n$-ality, as discussed in \cite{Tachikawa:2013hya},
\cite[section 2.4]{Sharpe:2014tca}).

Then, as discussed in \cite[section 2.4]{Sharpe:2014tca}),
decomposition becomes the statement
\begin{eqnarray}
Z(G) & = & \left( \frac{ {\rm Vol}(G) }{ {\rm Vol}(G/K) }
\right)^{2g-2} \sum_{\lambda \in \hat{K}} Z(G/K, \lambda),
\\
& = &
|K|^{2g-2}  \sum_{\lambda \in \hat{K}} Z(G/K, \lambda),
\label{eq:pureym:decomp:center}
\end{eqnarray}
Since $K$ is central and so abelian, in the language of~(\ref{eq:dilaton-shift-conj}),
every irreducible representation $R_U$ of $K$ necessarily has dimension one,
and so we see that the dilaton shift factor encoded by the Reidemeister-torsion-based
normalization, namely
\begin{equation}
|K|^{2g-2} \: = \: \left( \frac{1}{|K|} \right)^{2-2g},
\end{equation}
is of the form predicted by the dilaton shift conjecture~(\ref{eq:dilaton-shift-conj}).

\subsubsection{Pure Yang-Mills: decomposition to invertibles}
\label{sect:ym:inv}

In the previous subsections, we looked at decompositions of two-dimensional
pure gauge theories with trivially-acting subgroups $K$ for
$K$ finite, along the center $BK$ symmetry.
In this section we consider the decomposition in which $K$ is
no longer finite, and in fact $K=G$ because it is a pure gauge theory.
The resultig decomposition
pure Yang-Mills in two dimensions by noninvertible one-form
symmetries yields universes which are invertible field theories, as discussed in
\cite{Nguyen:2021yld,Nguyen:2021naa}.
We will see that the dilaton shift factors involve the
same factors of $(\dim R)^2$ that we have seen elsewhere.

Specifically, the proposal of \cite{Nguyen:2021yld,Nguyen:2021naa}
is that pure $G$ Yang-Mills in two dimensions decomposes into the
following union of invertible field theories, in the notation of
appendix~\ref{app:inv}:
\begin{equation}
\mbox{Pure $G$ Yang-Mills} \: = \: \coprod_R {\rm Inv}\left( - C_2(R),
\ln\left(\frac{(2\pi)^{\dim G}}{{\rm Vol}(G)} \dim R \right)\right),
\end{equation}
where the disjoint union is over irreducible representations $R$.
This is reflected in the fact that the partition function of
pure $G$ Yang-Mills on a Riemann surface $\Sigma$ of genus $g$ is
\cite{Migdal:1975zg,Drouffe:1978py,Lang:1981rj,Menotti:1981ry,Rusakov:1990rs,Witten:1991we,Thompson:1992hv,Blau:1991mp,Blau:1993hj,Cordes:1994fc,Witten:1992xu}
\begin{equation}
Z(G) \: = \: \left( \frac{ {\rm Vol}(G) }{ (2\pi)^{{\rm dim}\, G} } 
\right)^{2g-2} \sum_R \left( {\rm dim}\, R \right)^{2-2g}
\exp\left( - A C_2(R) \right),
\end{equation}
for $A$ the area of $\Sigma$, as discussed earlier.

Since this decomposition is derived by observing that the
entire gauge group $G$ acts trivially, and $|G|$ is infinite,
our previous expression for dilaton shifts~(\ref{eq:dilaton-shift-conj})
does not quite apply.  However, if we modify it, by taking the
universes to have dilatons
\begin{equation}
\left( \frac{ (2\pi)^{\dim K} }{ {\rm Vol}(K) } \dim R \right)^{2-2g}
\exp\left( - A C_2(R) \right)
\end{equation}
for $K=G$, instead of
\begin{equation}
\left( \frac{ \dim R_U }{|K|} \right)^{2-2g},
\end{equation}
then we get a close analogue of~(\ref{eq:dilaton-shift-conj}).

\subsubsection{Nonabelian $BF$ theory}

Now we turn to nonabelian $BF$ theory, or specifically
$BF$ theory for a nonabelian gauge group $G$, which for simplicity
we take to be connected and simply-connected, at level one.

Now, $BF$ theory is the zero-area limit of two-dimensional pure
Yang-Mills (see e.g.~\cite[section 2]{Witten:1991we}).
As that zero-area limit, it also has a decomposition to countably
many invertible field theories, indexed by irreducible representations of
the gauge group, just as pure Yang-Mills in two dimensions.

For simplicity, in this section we will assume the genus of the
Riemann surface $g > 1$.  (For smaller $g$, the exact expression for the
partition function as a series does not converge in the zero-area limit,
and must be regularized, see e.g.~\cite[section 2.5]{Blau:1993hj}.)

From the exact expression~(\ref{eq:2dym:part:exact}) for two-dimensional
pure Yang-Mills, we see that the
partition function of two-dimensional nonabelian $BF$ theory is
\begin{equation}  \label{eq:nonabelbf:part:exact}
Z(G) \: = \: \left( \frac{ {\rm Vol}(G) }{ (2\pi)^{{\rm dim}\, G} } 
\right)^{2g-2} \sum_R \left( {\rm dim}\, R \right)^{2-2g}
\end{equation}
on a Riemann surface of genus $g$.

This is in the expected
form
\begin{equation}
Z \: = \: \sum_R f(R)^{\chi}
\end{equation}
for some function $f(R)$.  

For connected and
simply-connected $G$, $BF$ theory at level one is equivalent to a disjoint
union
\begin{equation}  \label{eq:decomp-bf}
\coprod_R {\rm Inv}\left( 0, \ln \left( 
\frac{ (2 \pi)^{\dim G} (\dim R) }{ {\rm Vol}(G)} \right) \right)
\: \cong \:
\coprod_R {\rm Inv}\left( 0, \ln \left( \dim R \right) \right),
\end{equation}
where the disjoint union is over all irreducible representations of $G$.
As before, this matches the common form~(\ref{eq:common})
mentioned in the introduction (in the sense that the representation-dependence
is identical, omitting normalization constants).

$BF$ theory recently made an appearance in 
\cite{Kapec:2019ecr}, which reviewed the form of the partition function
above, 
as a special case of our general conjecture~(\ref{eq:dilaton-shift-conj}).

\subsubsection{Aside: moduli space volumes}
\label{sect:exs:gauge:volumes}

The results above for decomposition in
$BF$ theory are related to the symplectic volume
of the moduli space of flat connections.
We shall review results of  \cite{Witten:1991we,Witten:1992xu}
(see also e.g.~\cite{ho-jeff,km1,jk1,jparabolic,amwdh,amwv}), 
and describe their
understanding in terms of decomposition and
dilaton shift factors.

First, let us review some results of \cite{Witten:1992xu}.
Let ${\rm Vol}({\cal M},G)$ denote the symplectic volume of the moduli space of
flat $G$ connections over a fixed Riemann surface $\Sigma$.
From \cite[equ'n (4.19)]{Witten:1992xu}, the partition function $Z$
of $BF$ theory is related to the volume as 
\begin{eqnarray}  \label{eq:witten:vol:part}
Z(G) & = & \frac{ {\rm Vol}({\cal M},G) }{ |{\cal Z}(G)| },
\\
& = &
 \left( \frac{ {\rm Vol}(G) }{ (2 \pi)^{{\rm dim}\, G}} \right)^{2g-2}
\sum_R \left( \dim R \right)^{2-2g},      \label{eq:vol-sum}
\end{eqnarray}
where ${\cal Z}(G)$ denotes the center of $G$, and in the
second line, Vol$(G)$ denotes the volume of $G$ itself, rather than the
moduli space.

Note that decomposition of $BF$ theory into invertibles gives a simple
physical explanation for why the symplectic volume of the moduli space
of flat connections can be written in the form of a sum over
irreducible representations in~(\ref{eq:vol-sum}), 
and the fact that each irreducible
representation contributes a term proportional
to $(\dim R)^{\chi}$ is a consequence of the dilaton shift factors. 

Now, let us reconcile decomposition with
the results of \cite{Witten:1992xu} above,
by applying the analysis to 
two-dimensional $BF$ theories and pure Yang-Mills
with gauge group $G/K$, for $K$ a subgroup of the center,
with discrete theta angles $\lambda \in \hat{K}$.
Recall that the effect of adding a discrete theta angle is to weight
contributions to partition functions
defined by $G/K$ bundles with characteristic class
$w \in H^2(\Sigma,K)$ ($\Sigma$ the two-dimensional space) by
phase factors $\exp\left( \langle w, \lambda \rangle \right)$,
with $\langle, \rangle$ denoting the natural pairing.
Schematically, if $Z_w$ denotes the part of a partition function obtained
by summing over bundles of fixed characteristic class $w$, then
the whole partition function $Z(\lambda)$ for fixed discrete theta angle
$\lambda \in \hat{K}$ is
\begin{equation}
Z(\lambda) \: = \: \sum_{w \in H^2(\Sigma,K)} 
\exp\left( \langle w, \lambda \rangle \right) \, Z_w.
\end{equation}

The analysis in \cite[section 2]{Witten:1991we} relating
operator determinant ratios to the measure
defined by the symplectic form is local in nature, so we expect it to also
apply in this case to the individual components of the moduli space,
with the phase factors $\exp\left( \langle w, \lambda \rangle \right)$
just `going along for the ride.'
Thus, at least naively,
the analysis of \cite[section 2]{Witten:1991we}
seems to suggest that the partition function is related to a `weighted volume'
which adds volumes of different moduli space components (of fixed $w$)
weighted by
the same phase factors $\exp\left( \langle w, \lambda \rangle \right)$.

To that end, define Vol$({\cal M}, G/K, \lambda)$ to be the
weighted volume of
the moduli space of flat $G/K$ connections, $K$ a subset of the center of
$G$, $G$ simple and simply-connected, in which a component of
the moduli space with characteristic class $w \in H^2(\Sigma, K)$ is
weighted by the phase $\exp\left( \langle w, \lambda \rangle \right)$ for
$\lambda \in \hat{K}$.  Schematically, if $V_{w}$ is the volume of a 
component with fixed characteristic class $w$, then in other words,
\begin{equation}
{\rm Vol}(({\cal M}, G/K, \lambda) \: = \: \sum_{w \in H^2(\Sigma, K)}
\exp\left( \langle w, \lambda \rangle \right) \, V_{w}.
\end{equation}

Now, the partition function of two-dimensional $BF$ theory 
(the zero-area limit of pure Yang-Mills) with gauge group $G/K$
and discrete theta angle $\lambda$ is
 \cite{Tachikawa:2013hya}
\begin{equation}
Z(G/K,\lambda) \: = \:
 \left( 
\frac{ {\rm Vol}(G/K) }{(2\pi)^{\rm \dim G} } \right)^{2g-2}
\sum_{R:\lambda} \left( \dim R \right)^{2-2g},
\end{equation}
where the sum is over irreducible representations of $G$ (not necessarily
$G/K$) of $n$-ality $\lambda$, and where we have used the
normalization conventions of 
\cite[equ'n (4.19)]{Witten:1992xu}.
Applying~(\ref{eq:witten:vol:part})
suggests that
\begin{equation}  \label{eq:vol:weightedsum}
{\rm Vol}({\cal M}, G/K, \lambda) \: = \: 
| {\cal Z}(G/K) | \left( 
\frac{ {\rm Vol}(G/K) }{(2\pi)^{\rm \dim G} } \right)^{2g-2}
\sum_{R:\lambda} \left( \dim R \right)^{2-2g}.
\end{equation}
As before, the form of this expression is a consequence of the
decomposition of the $BF$ theory or pure Yang-Mills to invertibles,
and the terms $(\dim R)^{\chi}$ are a reflection of dilaton shifts.

Furthermore, decomposition (along the center $BK$ symmetry of
two-dimensional $BF$ or pure Yang-Mills) 
then makes a prediction relating weighted moduli
space volumes.  
Specifically, from the decomposition~(\ref{eq:pureym:decomp:center}),
namely
\begin{equation}
Z(G) \: = \: 
|K|^{2g-2}  \sum_{\lambda \in \hat{K}} Z(G/K, \lambda),
\end{equation}
and
the relation~(\ref{eq:witten:vol:part}) between partition functions and
moduli space volumes, we have
\begin{equation} \label{eq:vol:decomp}
\frac{ {\rm Vol}({\cal M}, G) }{ | {\cal Z}(G) | }
\: = \: 
|K|^{2g-2} \sum_{\lambda \in \hat{K}} 
\frac{ {\rm Vol}({\cal M}, G/K, \lambda) }{ | {\cal Z}(G/K) | },
\end{equation}
where ${\cal Z}(G)$ denotes the center of $G$.  If we make the
further simplifying assumption that $K = {\cal Z}(G)$ (so 
${\cal Z}(G/K) = \{1\}$), then
\begin{equation}
{\rm Vol}( {\cal M}, G) \: = \:
|K|^{2g-1} \sum_{\lambda \in \hat{K}} {\rm Vol}({\cal M}, G/K, \lambda).
\end{equation}

Next, we study the abstract statements above in a simple
concrete example, namely $G = SU(2)$ and $K={\mathbb Z}_2$.
In this case, we will be able to explicitly check the result above
for the weighted moduli space volume for $SO(3)$.

First, let us quickly review results of \cite{Witten:1991we} on
(unweighted) moduli space volumes for $SU(2)$ and $SO(3)$.
In the conventions of \cite[section 4.5]{Witten:1991we},
\begin{equation}  \label{eq:vol:su2}
{\rm Vol}(SU(2)) \: = \: 2^{5/2} \pi^2,
\end{equation}
hence, using~(\ref{eq:vol-sum}),
the symplectic volume Vol$({\cal M},SU(2))$ of the
moduli space of flat $SU(2)$ connections on a Riemann surface of genus $g$ is
\cite[equ'ns (3.11), (4.73)]{Witten:1991we}
\begin{equation}  \label{eq:vol:so3p}
{\rm Vol}({\cal M}, SU(2)) \: = \: 
\frac{2}{(2 \pi^2)^{g-1}} \sum_{n=1}^{\infty} n^{2-2g}
\: = \:
\frac{2}{(2 \pi^2)^{g-1}} \zeta(2g-2).
\end{equation} 
The values of $n$ above are simply the dimensions of irreducible
representations of $SU(2)$, corresponding to spins $j$, where $n=2j+1$.
Similarly, using the fact that in
the conventions of \cite[section 4.5]{Witten:1991we},
\begin{equation}
{\rm Vol}(SO(3)) \: = \: 2^{3/2} \pi^2,
\end{equation}
the volume of the moduli space of flat $SO(3)$ connections is
\cite[equ'n (3.29), (4.74)]{Witten:1991we}
\begin{equation}
{\rm Vol}({\cal M},SO(3)) \: = \: \frac{1}{(8 \pi^2)^{g-1}} \sum_{n=1, 3, 5, \cdots}
n^{-(2g-2)}.
\end{equation}
Here, $n$ indexes dimensions of irreducible representations of $SO(3)$.
To distinguish this quantity from the weighted moduli space volume,
and to follow notation for discrete theta angles,
we will use the notation
\begin{equation}
{\rm Vol}( {\cal M}, SO(3)_+) \: = \: {\rm Vol}({\cal M},SO(3)) .
\end{equation}
(Technically,
the $+$ subscript indicates that the corresponding QFT does not have
a discrete theta angle, to distinguish it from the next case we consider.)

Now, let us turn to weighted moduli space volumes.
Here, the characteristic class $w \in H^2(SO(3),{\mathbb Z}_2)$ is the
second Stiefel-Whitney class, so we will denote it $w_2$ in the remainder
of this section.
Let Vol$({\cal M},SO(3))_-)$ denote this weighted sum of
$SO(3)$ moduli space component volumes.  To make the description above
more concrete, define
\begin{itemize}
\item $V_0 = $ the sum of volumes of components of the moduli space of
flat $SO(3)$ connections with $w_2$ trivial, and
\item $V_1 = $ the sum of volumes of components of the moduli space of
flat $SO(3)$ connections with $w_2$ nontrivial.
\end{itemize}
The phase factors $\exp\left( \langle w, \lambda \rangle \right)$ simply
reduce to signs in this case.
Then,
\begin{eqnarray}
{\rm Vol}({\cal M}, SO(3)_+)
& = & V_0 + V_1, 
\label{eq:so3ex:1}
\\
{\rm Vol}({\cal M}, SO(3)_-)
& = & V_0 - V_1.
\label{eq:so3ex:2}
\end{eqnarray}

The partition function of the two-dimensional $SO(3)$ gauge theory with
nontrivial discrete theta angle is 
\cite{Tachikawa:2013hya}
\begin{equation}
Z( SO(3)_- ) \: = \:
 \frac{1}{(8 \pi^2)^{g-1}} \sum_{n=2,4,6,\cdots}
n^{-(2g-2)}.
\end{equation}
As the center of $SO(3)$ is trivial,
equation~(\ref{eq:vol:weightedsum}) reduces to
\begin{equation}
{\rm Vol}({\cal M},SO(3)_-) \: = \: Z(SO(3)_-)
 \: = \:  \frac{1}{(8 \pi^2)^{g-1}} \sum_{n=2,4,6,\cdots}
n^{-(2g-2)}.
\end{equation}
Here, $n$ indexes dimensions of allowed irreducible
representations -- of 
of $SU(2)$ representations that do not descend to $SO(3)$,
following \cite{Tachikawa:2013hya}.
This is easily checked to be true by algebraically solving
equations~(\ref{eq:so3ex:1}), (\ref{eq:so3ex:2}), verifying that
for $SO(3)$, our expression for weighted moduli space sums is correct.

Decomposition along the central $BK$ symmetry implied
equation~(\ref{eq:vol:decomp}), which in this case specializes to
\begin{eqnarray}
{\rm Vol}({\cal M},SO(3)_+) \: + \: {\rm Vol}({\cal M},SO(3)_-) & = &
 \frac{1}{(8 \pi^2)^{g-1}} \sum_{n=1}^{\infty} n^{-2g-2},
\\
& = &
 \frac{1}{(2) 4^{g-1} }  {\rm Vol}({\cal M},SU(2)),
\end{eqnarray}
or more simply
\begin{equation}
{\rm Vol}({\cal M},SU(2)) \: = \: (2) (2)^{2g-2} \left(
{\rm Vol}({\cal M},SO(3)_+) \: + \: {\rm Vol}({\cal M},SO(3)_-) \right).
\end{equation}
This is also easily checked to be true just from the existing
results~(\ref{eq:vol:su2}), (\ref{eq:vol:so3p}).

For $SO(3)$, the single weighted moduli space volume Vol(${\cal M}, SO(3)_-$)
can be deduced using the fact that the corresponding volumes for both
$SU(2)$ and $SO(3)_+$ were known.  For higher-rank cases, there can be
additional weighted moduli space volumes (correspoinding to
all of the $\lambda \in \hat{K}$), so knowledge of moduli space
volumes for just $G$, $G/K$ alone do not always suffice to algebraically
determine the remainder.

\subsection{Examples in TFTs}
\label{sect:exs:tfts}

Further examples of decomposition are provided by unitary
topological field theories with semisimple local operator algebras,
such as two-dimensional Dijkgraaf-Witten theory, $BF$ theory,
and the $G/G$ model.  
We have already examined some of these theories; in this section,
we will provide a systematic construction as topological field theories,
and examine those theories as special cases of that construction.

Existence of a decomposition of these theories
is a consequence of the fact, well-known in the TFT community,
that the operator algebras of such TFTs admit a complete set of 
projectors.  
This implies that these theories are equivalent to
disjoint unions of invertible field theories,
which is a special case\footnote{
Decompositions go hand-in-hand with global one-form symmetries in
two-dimensional theories.  For unitary semisimple TFTs, the one-form
symmetries responsible for their description as a decomposition
are described in e.g.~\cite{Gukov:2021swm,Huang:2021zvu}.
} of decomposition, see
e.g.~\cite{Durhuus:1993cq}, \cite[section 3.1]{Moore:2006dw},
and more recently,
\cite[appendix C.1]{Komargodski:2020mxz}, \cite{Huang:2021zvu}.

We will see that in such two-dimensional topological field theories,
the partition function has the form
\begin{equation}
\sum_U \left( S_{00} \dim_q \Pi_U \right)^{2-2g},
\end{equation}
where here dim$_q$ indicates the quantum dimension.
The fact that the dilaton shift on
each universe is proportional to $( \dim_q \Pi_U )^{2-2g}$ is closely comparable
to results discussed in previous sections.

We will compute this in examples and compare to previous results where
applicable.

In subsection~\ref{sect:tft:genl}, we begin by giving a general
analysis of the decomposition and dilaton shifts of two-dimensional
unitary topological field theories.  We then work out the details in
two examples.  In subsection~\ref{sect:tft:dw}, we return to
two-dimensional Dijkgraaf-Witten theory, which we now view as an
example of a unitary topological field theory, and recover the previous
description of dilaton shifts from sections~\ref{sect:2ddw:1}, 
\ref{sect:2ddw:2} as a special
case of general aspects of topological field theories.
Then, in subsection~\ref{sect:tft:gg}, we consider $G/G$ models,
and analyze their decompositions and dilaton shifts in examples.

Other examples of unitary topological field theories also exist,
including nonabelian $BF$ theory and abelian $BF$ theory at various
levels.  These also decompose, as has been discussed elsewhere
(see the zero-area limit of \cite{Nguyen:2021yld,Nguyen:2021naa} 
for nonabelian $BF$ theory, \cite{Hellerman:2010fv} for
abelian $BF$ theory), but for the purpose of outlining topological
field theories in this framework, Dijkgraaf-Witten theory and the $G/G$
models will suffice.

We hasten to add that we have deliberately restricted to unitary TFTs,
meaning we exclude cohomological field theories (obtained by a topological
twist of a supersymmetric theory).  In these theories, the topological
subsector may also admit a decomposition, but the topological
subsector is only one small slice of a much larger QFT, and typically
in those examples, the full QFT does not decompose, only the topological
subsector.

In passing, the fact that unitary TFTs decompose into invertible field theories
was utilized recently in studies of factorization issues in
AdS/CFT, see e.g.~\cite{Marolf:2020xie,Gardiner:2020vjp,Banerjee:2022pmw,deMelloKoch:2021lqp,Benini:2022hzx}.

\subsubsection{General remarks}
\label{sect:tft:genl}

We can describe dilaton shifts in the decomposition of unitary 
two-dimensionsl TFTs in a simple general fashion,
as we now review, following e.g.~\cite{Dijkgraaf:1990qw,Fukuma:1993hy,Karimipour:1995fb},
\cite[appendix C]{Komargodski:2020mxz}.

Briefly, the local operators form a commutative Frobenius algebra $F$,
meaning (among other things) that there is a linear trace map
$\theta: F \rightarrow {\mathbb C}$.  This trace map defines the
metric and correlation functions.  For example, if ${\cal O}_i$ denotes
local operators in $F$, then
correlation functions on $S^2$ are
\begin{equation}
\langle {\cal O}_1 \cdots {\cal O}_n \rangle_0 \: = \:
\theta\left( {\cal O}_1 \cdots {\cal O}_n \right),
\end{equation}
where the subscript $0$ emphasizes that this is a correlation function
in genus zero.
Similarly,
the topological metric is defined to be a two-point function on $S^2$
\begin{equation}
g_{ij} \: = \: \langle {\cal O}_i {\cal O}_j \rangle_0 \: = \:
\theta\left( {\cal O}_i {\cal O}_j \right).
\end{equation}

Let $C_{ijk}$ denote the three-point function on $S^2$
\begin{equation}
C_{ijk} \: = \: \langle {\cal O}_i {\cal O}_j {\cal O}_k \rangle_0
\end{equation}
or, equivalently,
\begin{equation}
{\cal O}_i \cdot {\cal O}_j \: = \: \sum_k C_{ij}^{\: \: \: k} \, {\cal O}_k,
\end{equation}
and define the handle-attaching operator $H: F \rightarrow F$ by
(see e.g.~\cite[section 4.1]{Nekrasov:2014xaa})
\begin{equation}
H \: = \: \eta^{ij} C_{ij}^{\: \: \: k} \, {\cal O}_k,
\end{equation}
or in components,
\begin{equation}
H^i_j \: = \: C^{ik\ell} C_{k \ell j},
\end{equation}
where the indices are raised with the topological metric above.
Then, the partition function on a genus $g$ surface is
\begin{equation}  \label{eq:tftz}
Z(\Sigma_g) \: = \: \langle H^g \rangle_0 \: = \: \theta\left( H^g  \right).
\end{equation}

Semisimplicity implies that there is a basis $\{ \Pi_i \}$ of
the local operators such that
\begin{equation}
\Pi_i \Pi_j \: = \: \delta_{ij} \Pi_i,
\: \: \:
\sum_i \Pi_i \: = \: 1.
\end{equation}
We will refer to the elements of this basis as projectors,
for obvious reasons.
Unitarity requires that the one-point function
\begin{equation}
\theta_i \: = \: \langle \Pi_i \rangle_0 \: = \: \theta\left( \Pi_i \right)
\end{equation}
be a positive real number \cite[appendix C.1]{Komargodski:2020mxz}.

Now, let us compute the partition function from the expression~(\ref{eq:tftz}).
If we work in a basis of projectors, then the topological metric
is diagonal, and we write
\begin{equation}
g_{ij} \: = \: \langle \Pi_i \Pi_j \rangle_0 \: = \: \langle \Pi_i \rangle_0
\, \delta_{ij} \: = \: \theta_i \, \delta_{ij}.
\end{equation}
Similarly,
\begin{equation}
C_{ijk} \: = \: \left\{ \begin{array}{cl}
\theta_i & i=j=k,
\\
0 & {\rm else}.
\end{array}
\right.
\end{equation}
As a result, the handle-attaching operator is
\begin{equation}
H \: = \: \sum_i  \left( \theta_i \right)^{-1} \Pi_i,
\end{equation}
or in components,
\begin{equation}
H^i_j \: = \: \left( \theta_i \right)^{-1} \delta_{ij}.
\end{equation}
A correlation function on a genus $g$ Riemann surface is then
\begin{equation}
\langle {\cal O}_1 \cdots {\cal O}_n \rangle_g \: = \:
\theta \left( {\cal O}_1 \cdots {\cal O}_n H^g \right),
\end{equation}
and more pertinently, the partition function is
(see e.g.~\cite[equ'n (C.4)]{Komargodski:2020mxz})
\begin{equation}  \label{eq:2dtft:partfn}
Z(\Sigma_g) \: = \:
\langle  H^g \rangle_0 \: = \: 
\sum_i \left( \theta_i \right)^{1-g}.
\end{equation}
Then, the dilaton shift in the $i$th universe is 
given by
\begin{equation}
\frac{1}{2} \ln \theta_i,
\end{equation}
where
$\theta_i = \langle \Pi_i \rangle$.
(As in section~\ref{sect:orb:nodt:conj}, this may be more invariantly
understood as a shift relative to that of the ambient theory.)

To understand this more elegantly, let us briefly review the role of
the modular S-matrix.  (In the remainder of this section, we will also
restrict to TCFTs (topological conformal field theories), for simplicity.)

In conventions in which ${\cal O}_0  = 1$,
the fusion rules
are (famously in RCFT) diagonalized by the modular $S$-matrix
(see e.g.~\cite[equ'n (3.11)]{Verlinde:1988sn},
\cite[equ'n (9.57)]{Ginsparg:1988ui}, \cite{Moore:1988uz,Dijkgraaf:1988tf})
\begin{equation}
{\cal O}_i {\cal O}_j
 \: = \:
 \sum_{mn} S_{im} S_{jm} \left( \frac{ (S^{\dag})_{mn} }{ S_{0m} }
\right)
{\cal O}_n.
\end{equation}
Given the fusion rules above, one can define projectors
\cite[equ'n (C.16)]{Komargodski:2020mxz}
\begin{equation}  \label{eq:proj-genl}
\Pi_i \: = \: S_{0i} \sum_p (S^{\dag})_{ip} \, {\cal O}_p,
\end{equation}
and it is straightforward to check that
\begin{equation}
\Pi_i \Pi_j \: = \: \delta_{ij} \Pi_i, \: \: \:
\sum_i \Pi_i \: = \: 1,
\end{equation}
which are both a consequence of the unitarity of $S$,
$S^{\dag} = S^{-1}$.

Note that conversely,
given the projectors above, we can compute the fusion rules.
First, inverting~(\ref{eq:proj-genl}), one quickly finds
\begin{equation}
{\cal O}_p \: = \: \sum_m S_{pm} \left( \frac{\Pi_m}{S_{0m} } \right),
\end{equation}
which implies
\begin{eqnarray}
{\cal O}_p {\cal O}_q & = & \left( \sum_i S_{pi} \frac{ \Pi_i}{S_{0i}} \right)
\left( \sum_j S_{pj} \frac{\Pi_j}{S_{0j}} \right),
\\
& = &
\sum_m S_{pm} S_{qm} \frac{\Pi_m}{ (S_{0m})^2 },
\\
& = &
\sum_m S_{pm} S_{qm} \sum_n \frac{ (S^{\dag})_{mn} }{ S_{0m} } {\cal O}_n,
\end{eqnarray}
giving another perspective on the diagonalization of the fusion rules.

The partition function at genus $g$ is then
(see e.g.~\cite[equ'n (C.4)]{Komargodski:2020mxz})
\begin{equation}   \label{eq:tft:part}
Z_g \: = \: \sum_{i} \left( S_{0 i} \right)^{2-2g},
\end{equation}
where
$S_{0 i}$ is the matrix element for $0$ denoting the operator corresponding
to the identity.  
From \cite[appendix C.2]{Komargodski:2020mxz}, in terms of the $S$
matrix above, the quantum dimension dim$_q$ of the operator ${\cal O}_i$ is
\begin{equation}  \label{eq:qdim:tft}
\dim_q {\cal O}_i \: = \: \frac{ S_{0i} }{ S_{00} },
\end{equation}
hence in terms of the quantum dimension dim$_q$, we see that
\begin{equation}
Z_g \: = \: \sum_i \left( S_{00} \dim_q {\cal O}_i \right)^{2-2g}.
\end{equation}
Thus, we see that the contribution from the $i$th universe is proportional to
$(\dim_q \Pi_i)^{2-2g}$, hence the
dilaton shift in the $i$th universe is proportional to
$\ln \dim_q \Pi_i$.

\subsubsection{Two-dimensional Dijkgraaf-Witten theory}
\label{sect:tft:dw}

Previously in section~\ref{sect:2ddw:1} we discussed decomposition in
two-dimensional Dijkgraaf-Witten theories, viewed as orbifolds of points.
In this section we return to Dijkgraaf-Witten theories, now viewing
them as examples of unitary topological field theories, in which to
apply the abstract machinery described above.

First, we note that the projector onto a universe corresponding
to irreducible (projective) representation $R$ is given by
\cite[equ'n (2.17)]{Ramgoolam:2022xfk},
\cite[section 7.3]{karpilovsky}
\begin{equation}  \label{eq:dw:projector}
\Pi_R \: = \:
\frac{\dim R}{|G|} \sum_{g \in G} \frac{ \chi_R(g^{-1}) }{ 
\omega(g,g^{-1}) } \, g,
\end{equation}
where $[\omega] \in H^2(G,U(1))$ denotes the choice of discrete torsion
(normalized so that $\omega(1,g) = \omega(g,1) = 1$)
in the $G$ orbifold.

Now, correlation functions have the form (see e.g.~\cite{Ramgoolam:2022xfk})
\begin{equation}
\langle U_1 \cdots U_b \rangle \: = \:
\sum_R \left( \frac{\dim R}{|G|} \right)^{\chi(\Sigma)}
\frac{\chi_R(U_1)}{\dim R} \cdots \frac{\chi_R(U_b)}{\dim R},
\end{equation}
for $U_i$ elements of the center of the group algebra ${\mathbb C}[G]$
(which forms the space of operators in the Dijkgraaf-Witten theory),
and
\begin{eqnarray}
\chi_R(\Pi_S) & = &
\frac{\dim S}{|G|} \sum_{g \in G}
 \frac{ \chi_S(g^{-1}) }{ 
\omega(g,g^{-1}) } \chi_R(g),
\\
& = &
\delta_{R,S} \dim R,
\end{eqnarray}
using identity~(\ref{eq:char-om1}).

As a result,
\begin{equation}
\langle \Pi_R \rangle \: = \: \left( 
\frac{\dim R}{|G|} \right)^{2-2g}
\end{equation}
on a Riemann surface $\Sigma$ of genus $g$.

In the abstract language of subsection~\ref{sect:tft:genl},
we reviewed how the partition function of any unitary semisimple TFT should
have the form
\begin{equation}
Z \: = \: \sum_R \left( \theta_R \right)^{1-g}
\end{equation}
for $\theta_R = \langle \Pi_R \rangle_0$ the expectation value of the
projector on $S^2$.
Here, we can confirm that explicitly, as the partition function of
two-dimensional Dijkgraaf-Witten theory on a surface of genus $g$ is
\begin{equation}
Z(\Sigma_g) \: = \:
\sum_R
\left( 
\frac{\dim R}{|G|} \right)^{2-2g}
 \: = \:
\sum_R \left( \theta_R \right)^{1-g}
\end{equation}
for
\begin{equation}
\theta_R  \: = \: \langle \Pi_R \rangle_0 \: = \:
\left( 
\frac{\dim R}{|G|} \right)^{2}
\end{equation}
(the expectation value at genus zero).
This matches the earlier description~(\ref{eq:2ddw:part:1}).
There, we viewed Dijkgraaf-Witten theory as an orbifold of a point;
here, we have approached it abstractly as a topological field theory.
Just as there, the factor
\begin{equation}
\left( 
\frac{\dim R}{|G|} \right)^{2-2g}
\end{equation} 
is interpreted in the decomposition in terms of a dilaton shift.

In the special case of vanishing discrete torsion,
$S$ matrix elements can be written explicitly for
two-dimensional Dijkgraaf-Witten theory,
following \cite[exercise 10.18]{diFranc}.
(We will recover the partition function up to an overall $R$-independent
constant factor.)
For a finite group $G$ with representation $R$ and character $\chi_R$,
(and no discrete torsion,)
one can define a group $S$-matrix on any conjugacy class $[g]$
as
\cite[equ'n (A.8)]{Dijkgraaf:1989hb},
\cite[equ'n (10.277)]{diFranc} (see also \cite[section 2.2]{Coste:2000tq})
\begin{equation}  \label{eq:2ddw:s}
S_R([g]) \: = \: \left( \frac{ | [g] | }{ |G| } \right)^{1/2}
\chi_R(g)
\end{equation}
that diagonalizes products of irreducible representations, meaning
\cite[equ'n (10.276)]{diFranc}
\begin{eqnarray}
C_{RS}^{ \: \: \: T} & = & \frac{1}{|G|} \sum_{[g]} |[g]| \chi_R(g) \chi_S(g)
\overline{\chi}_T(g),
\\
& = &
\sum_{[g]} S_R(g) \frac{ S_S(g) }{ S_1(g) } \overline{S}_T(g).
\end{eqnarray}
for $R$, $S$, $T$ irreducible representations.
In particular, in this fashion we can recover the Dijkgraaf-Witten
projectors, as described in the general analysis of
section~\ref{sect:tft:genl}.
Following~(\ref{eq:proj-genl}),   
and letting ${\cal O}_{[g]}$ denote the twist field associated to
conjugacy class $[g]$, 
normalized as
\begin{equation}
{\cal O}_{[g]} \: = \: \frac{1}{ | [g] |^{1/2} } \sum_{h \in [g]} h,
\end{equation}
we have
\begin{eqnarray}
\Pi_R & = &
S_{0R} \sum_{[g]} (S^{\dag})_{R}( [g]) \, {\cal O}_{[g]},
\\
& = &
 \chi_R(1)   \sum_{[g]} \frac{ | [g] |^{1/2} }{ |G| } \,
\overline{\chi}_R(g) \, {\cal O}_{[g]},
\\
& = &
\frac{\dim R}{|G|} \sum_{[g]} \chi_R(g^{-1}) \, |[g]|^{1/2} \, {\cal O}_{[g]},
\\
& = &
\frac{\dim R}{|G|} \sum_{g \in G} \chi_R(g^{-1}) \, g,
\end{eqnarray}
where in the last line we have
accounted for the multiplicity in conjugacy class elements.
This matches the Dijkgraaf-Witten projectors~(\ref{eq:dw:projector})
(for vanishing discrete torsion).

In terms of quantum dimensions, recall from~(\ref{eq:qdim:tft}) that
the quantum dimension of the projector $\Pi_R$ is given by
\begin{equation}
\dim_q \Pi_R \: = \: \frac{ S_{0R} }{ S_{00} }.
\end{equation}
From the definition~(\ref{eq:2ddw:s}),
\begin{equation}
S_{0R} \: = \: S_R(1) \: = \: \frac{1}{|G|^{1/2}} \, \dim R,
\: \: \:
S_{00} \: = \:  \frac{1}{|G|^{1/2}} \chi_1(1) \: = \:  \frac{1}{|G|^{1/2}} ,
\end{equation}
hence
\begin{equation}
\dim_q \Pi_R \: = \: \dim R,
\end{equation}
and the predicted partition function for the universe $R$ is
\begin{equation}
\left( S_{0R} \right)^{2-2g} \: = \:
\left( \frac{ \dim R}{ |G|^{1/2} } \right)^{2-2g},
\end{equation}
which matches previous results up to an overall convention-dependent
$R$-independent factor
involving $|G|$.

We only discuss
$S$ matrices for two-dimensional Dijkgraaf-Witten theory
without discrete torsion,
because 
the
analogue of $C^{\: \: \: k}_{ij}$
for projective representations is more complicated, due to the fact that
the tensor product of projective representatives for a fixed twist
does not close onto itself.  (The product of an
$\alpha$-twisted representation and a $\beta$-twisted representation is
an $\alpha \beta$-twisted representation.)

\subsubsection{$G/G$ model}
\label{sect:tft:gg}

The $G/G$ model is a (bosonic\footnote{
One can supersymmetrize gauged WZW models and topologically twist.
However, that is not our intent here -- we are describing
ordinary bosonic gauged WZW models, without fermions.
}) gauged WZW model $G/H$ (see e.g.~\cite{Witten:1991mk})
for the special
case that $H=G$.  In this special case, the gauged WZW model
is a topological field theory (see e.g.~\cite[section 4]{Witten:1991mm}).
This is a standard example in two dimensions,
with relations to other two-dimensional topological gauge theories.
We outline in this section
its decomposition to invertible field theories as a unitary
topological field theory, as in section~\ref{sect:tft:genl},
and discuss dilaton shifts.

The $G/G$ model is
discussed in detail in
e.g.~\cite{Witten:1991mm,Spiegelglas:1992jg,Spiegelglas:1989xp,Komargodski:2020mxz}, to which
we refer the reader.  As described there, the physical states
of the $G/G$ model at level $k$ correspond to conformal primaries of
the $G$ WZW model at level $k$, i.e., integrable representations of the
corresponding Kac-Moody algebra, with OPEs corresponding to
fusion rules in the same WZW model.

The physical states all have dimension zero
\cite[section 4]{Spiegelglas:1992jg}.
Since there exist multiple dimension-zero states, one expects
a decomposition, and since they are all dimension-zero, one expects
a decomposition to a disjoint union of invertible field theories,
whose form we shall outline momentarily.

Since the states are
dimension-zero, one can construct a complete
set of projection operators, which can be done using the
modular $S$ matrix from the general formula~(\ref{eq:proj-genl})
for any two-dimensional topological field theory. 
We will apply this in examples later.

For $G$ connected and simply-connected, the partition function of the
$G/G$ model at level $k$ equals the dimension of the corresponding
Chern-Simons Hilbert space
(see e.g.~\cite[section 3.4]{Blau:1993hj}),
which at genus $g$ is
\cite[equ'n (3.15)]{Verlinde:1988sn}, \cite{gukovprivate},
\begin{equation}
Z_g \: = \: \sum_{i} \left( S_{0 i} \right)^{2-2g},
\end{equation}
the same result described earlier in equation~(\ref{eq:tft:part}) for
any two-dimensional unitary topological field theory,
where
$S_{0 i}$ is proportional to the quantum dimension
of the integrable representation $i$, and
the sum is again over integrable representations $i$
of the Kac-Moody
algebra at level $k$.

As a result,
the
$G/G$ model at level $k$ decomposes into (is equivalent to)
the following disjoint union of invertible field theories
(see e.g.~\cite[section C.1]{Komargodski:2020mxz}):
\begin{equation}  \label{eq:decomp-gg}
\left( G/G \right)_k \: = \: \coprod_{i}
{\rm Inv}\left( 0, \ln S_{0 i} \right)
\: \cong \:
\coprod_{i}
{\rm Inv}\left( 0, \ln \left( \dim_q R_i \right) \right),
\end{equation}
indexed by the integrable representations of the $G$
Kac-Moody algebra at level $k$, and where
dim$_q R_i$ denotes the quantum dimension, $S_{0i}/S_{00}$.
This matches the common form~(\ref{eq:common}) described in the 
general analysis of section~\ref{sect:tft:genl}.

In passing, in addition to local operators, the $G/G$ model also contains
`Verlinde' line operators $L_p$, dimensional reductions of Wilson lines in
three-dimensional Chern-Simons theory, which obey the same fusion relations
as the local operators $x_p$ above.  As a result, one can form an identical
projector from the Verlinde lines,
\begin{equation}
\Pi^L_i \: = \: S_{0i} \sum_p \left( S^{\dag} \right)_{ip} L_p,
\end{equation}
which acts as a (nondynamical) domain wall separating universes.

Next, for completeness, we outline two examples.
First, we consider
$SU(n)$ algebras.
From \cite[section 8.3]{Distler:2007av},
\begin{itemize}
\item $SU(n)$ integrable reps at level 1 are antisymmetric powers of the
fundamental,
\item $SU(n)$ integrable reps at level $k$ are Young diagrams of width
bounded by the level.
\end{itemize}
For example, at level 2, the adjoint representation of $SU(n)$ becomes
integrable.

For integrable $SU(n)$ representations at level $1$,
the fusion algebra is
\cite[equ'n (3.7)]{Walton:1989sc}
\begin{equation}
\left[ \wedge^i {\bf n} \right] \times \left[ \wedge^j {\bf n} \right]
\: = \: \left[ \wedge^{i + j \mod n} {\bf n} \right],
\end{equation}
which can be expressed more compactly as the ring
\begin{equation}
{\bf C}[x_1, \cdots, x_{n-1}] / \left( x_i x_j - x_{i + j \mod n} \right).
\end{equation}
For $n > 2$, there are more than $n-1$ constraints, so generically
one might expect no solutions, but it is straightforward to check that
solutions always exist of the form
\begin{equation}
x_1^n \: = \: 1, \: \: \: x_i \: = \: (x_1)^i,
\end{equation}
and so describe $n$ points.

To compare to the modular $S$-matrix, let us specialize to
$SU(3)_1$.
The integrable representations are
${\bf 1}$, ${\bf 3}$, $\overline{\bf 3} = \wedge^2 {\bf 3}$, and the
fusion rules are
\begin{equation}
\left[ {\bf 3} \right] \times [{\bf 3}] \: = \: \overline{\bf 3},
\: \: \:
[{\bf 3}] \times [\overline{\bf 3}] \: = \: [{\bf 1}],
\: \: \:
[\overline{\bf 3}] \times [\overline{\bf 3}] \: = \: [ {\bf 3}].
\end{equation}
If we identify $x_1$ with $[{\bf 3}]$ and $x_2$ with
$[\overline{\bf 3}]$, then the fusion algebra is the ring
\begin{equation}
{\mathbb C}[x_1, x_2] / \left( x_1^2 - x_2, x_1 x_2 - 1, x_2^2 - x_1 \right).
\end{equation}
This is a system of three equations in two unknowns -- an overdetermined
system.  Nevertheless, it does admit solutions, corresponding to a set
of three points, located at
\begin{equation}
x_1^3 \: = \: 1, \: \: \: x_2 = x_1^{-1}.
\end{equation}

The fusion ring for $SU(3)_1$ is encoded in the modular $S$-matrix
\cite[equ'n (14.222)]{diFranc}
\begin{equation}
        S \: = \: \frac{1}{\sqrt{3}} \left[ \begin{array}{ccc}
                1 & 1 & 1 \\
                1 & \xi & \xi^2 \\
                1 & \xi^2 & \xi 
        \end{array} \right],
\end{equation}
for $\xi = \exp(2 \pi i/3)$.
(Note that $S^{\dag} = S^{-1}$, as expected for a unitary matrix.)

Given the modular $S$-matrix, from the general formula~(\ref{eq:proj-genl})
we have
the projection operators
\begin{eqnarray}
\Pi_1 & = & \frac{1}{3} \left( 1 + x_1 + x_2 \right),
\\
\Pi_2 & = & \frac{1}{3} \left( 1 + \xi^2 x_1 + \xi x_2 \right),
\\
\Pi_3 & = & \frac{1}{3} \left( 1 + \xi x_1 + \xi^2 x_2 \right),
\end{eqnarray}
(which in this case can also be easily computed directly,
without knowledge of the
$S$-matrix).
These are easily checked to obey
\begin{equation}
\Pi_i \Pi_j \: = \: \delta_{ij} \Pi_i,
\: \: \:
\sum_i \Pi_i \: = \: 1,
\end{equation}
as expected for projection operators.

From equation~(\ref{eq:tft:part}), the partition function is then
\begin{equation}
Z_g \: = \: 
\sum_i \left( S_{0i} \right)^{2-2g} \: = \:
\sum_i \left( \frac{1}{\sqrt{3}} \right)^{2-2g},
\end{equation}
hence each universe is weighted by a factor of
\begin{equation}
 \left( \frac{1}{\sqrt{3}} \right)^{2-2g},
\end{equation}
reflecting a dilaton shift of 
\begin{equation}
- \frac{1}{2}\ln 3.
\end{equation}
Furthermore, from~(\ref{eq:qdim:tft}),
the quantum dimension of each projector is
\begin{equation}
\dim_q \Pi_i \: = \: \frac{S_{0i}}{S_{00}} \: = \: 1.
\end{equation}

Next, we outline the example of $G_2/G_2$ at level one.
From \cite[section 6]{Distler:2007av},
\cite[section 3]{Walton:1989sc},
there are two integrable representations of $G_2$ at level 1,
namely $[{\bf 1}]$, $[{\bf 7}]$, which obey
\begin{equation}
\left[{\bf 7}\right] \times [{\bf 7}] \: = \:
[{\bf 1}] \: + \: [{\bf 7}].
\end{equation}
We can write the ring as
\begin{equation}
{\mathbb C}[x] / (x^2 - 1 - x),
\end{equation}
where we have identified $x$ with $[{\bf 7}]$.
(In passing, $F_4$ at level 1 also has only two integrable representations,
which obey a fusion algebra of the same form.)
Geometrically, this ring describes two points, located at
\begin{equation}
x \: = \: \frac{1 \pm \sqrt{5}}{2},
\end{equation}
corresponding to the two universes (and to the roots of the quadratic
polynomial $x^2 - x - 1$).
The $S$ matrix is \cite[equ'n (16.64)]{diFranc}
\begin{equation}  \label{eq:g2:smatrix}
S \: = \: \sqrt{ \frac{4}{5} } \left[ \begin{array}{cc}
\sin (\pi/5) & \sin (3 \pi/5) \\
\sin (3 \pi/5)  & - \sin (\pi/5)
\end{array}
\right].
\end{equation}
It is straightforward\footnote{
The following may be helpful:
\begin{equation}
\sin (\pi/5) \: = \: \frac{ \sqrt{10 - 2 \sqrt{5} } }{4},
\: \: \:
\sin (3 \pi/5) \: = \: \frac{ \sqrt{10 + 2 \sqrt{5}} }{4}.
\end{equation}
} to check from the general
formula~(\ref{eq:proj-genl}) (or directly) that the
two projectors are
\begin{equation}
\Pi_{\pm} \: = \: 
\frac{1}{2}\left( 1 \pm \frac{ \sqrt{5} }{5} \right) \mp 
\frac{ \sqrt{5}}{5} x.
\end{equation}

From the $S$ matrix~(\ref{eq:g2:smatrix}),
From the general expression for partition functions in topological
field theories~(\ref{eq:tft:part}), we have that
\begin{equation}
Z_g \: = \: \sum_i \left( S_{0i} \right)^{2-2g}
\: = \:
\left( \sqrt{\frac{4}{5}} \sin(\pi/5) \right)^{2-2g} \: + \:
\left( \sqrt{\frac{4}{5}} \sin( 3 \pi/5) \right)^{2-2g},
\end{equation}
which is interpreted to mean that
one universe is weighted by
\begin{equation}
\left( \sqrt{\frac{4}{5}} \sin(\pi/5) \right)^{2-2g}
\end{equation}
and the other by
\begin{equation}
\left( \sqrt{\frac{4}{5}} \sin( 3 \pi/5) \right)^{2-2g},
\end{equation}
with corresponding dilaton shifts.
We can also read off from equation~(\ref{eq:qdim:tft}) that the quantum
dimensions of the two projectors are
\begin{equation}
1, \: \: \:
\frac{\sin(3\pi/5)}{\sin(\pi/5)}.
\end{equation}

Finally, let us relate the $G/G$ model to other models discussed
in this overview.
First, we have already provided $S$ matrix elements for two-dimensional
Dijkgraaf-Witten theory that enable it to be treated in a fashion closely
related to the $G/G$ model.  Also,
it is believed that in the limit of large level,
the $G/G$ model reduces to $BF$ theory
(see e.g.~\cite[section 3.3]{Blau:1993hj}),
which implies that in that limit, the decomposition of the $G/G$ model
should become the decomposition of $BF$ theory.  To that end, we note
that in the limit of large level, all representations become integrable,
and the quantum dimensions of the integrable representations become the
ordinary dimensions (see e.g.~\cite[section 16.3]{diFranc}).
Thus, the decomposition of the $G/G$ model into invertibles given
by~(\ref{eq:decomp-gg}) reduces to the decomposition of $BF$ theory,
up to an irrelevant overall dilaton shift,
as expected.

In passing, we have only mentioned $G/G$ cosets, but there exist
more general $G/H$ cosets.  Since $H$ acts by adjoints, if $H$ has
a center $Z(H)$ then the gauged $G/H$ WZW model has an $Z(H)$ one-form
symmetry, and so decomposes, as is discussed in examples in
\cite{Komargodski:2020mxz}.

\section{Noninvertible symmetries and asymptotic densities of states}
\label{sect:consistency-asymp}

So far in this paper we have discussed dilaton shift (Euler counterterm)
factors arising in the universes of decomposition, and how they have
a more or less canonical form, proportional to the dimension of the
representation corresponding to the universe (with other factors that
are convention-dependent).

That same dim $R$ dependence was also recently discussed in
\cite{Bhardwaj:2023idu}, in the context of gapped theories,
which in the IR can be thought of as special cases of decomposition.
Briefly, they argue in those special cases
that the fact that the dilaton shifts (Euler counterterms) have a canonical
form is due to the presence of a (noninvertible) symmetry,
see e.g.~\cite[section 2]{Bhardwaj:2023idu}.
More precisely, they argue that due to linking between
one-dimensional interfaces between vacua, there are relative Euler terms,
essentially arising as the quantum dimensions of those interfaces,
see in particular \cite[section 2.3]{Bhardwaj:2023idu}.
One-dimensional interfaces also exist more generally
between universes, see e.g.~\cite{Sharpe:2019ddn,Sharpe:2021srf},
and using linking between those interfaces and the (local) univese
projection operators, the same argument applies and
one immediately reaches the same conclusion,
that there is a relative Euler counterterm shift between contributions
from different universes, with Euler counterterm proportional to 
\begin{equation}
\ln \left( \frac{ \dim R_i }{\dim R_j } \right),
\end{equation}
where $R_{i,j}$ are the representations associated with either
universe. 
Similar ideas also appear in
\cite{Vandermeulen:2023zec}.

We note for our purposes in this paper that 
in an orbifold or gauge theory in which a subgroup $K$ acts trivially,
there is a Rep($K$) quantum symmetry \cite{Bhardwaj:2017xup}, 
(possibly a subset of a larger quantum symmetry,)
which will be noninvertible if
$K$ is nonabelian.  We therefore interpret the form of these
dilaton shifts in gauge theories\footnote{
We have discussed a variety of theories, and it is not c
} in terms of the presence of such symmetries in the
decomposition.

The remarks above are meant to be specific to
gauge theories.  For example, two-dimensional unitary topological
field theories also decompose \cite{Durhuus:1993cq,Moore:2006dw}, 
but we are not aware of
relevant corresponding symmetries applicable to all such.

In passing, related notions have arisen in discussion of asymptotic
state densities, see e.g.~\cite{Cardy:1986ie,Pal:2020wwd,Lin:2022dhv,Kapec:2019ecr}.
For example, \cite[section 4]{Kapec:2019ecr} related asymptotic state densities
on boundaries of two-dimensional $G$ gauge theories
to $(\dim R)^2$ factors in bulk partition functions.
In such a case, the bulk theory has a global Rep$(G)$ symmetry,
but the boundary theory has a global $G$ symmetry.

\section{General argument via coupling to a TFT}
\label{sect:tftcoupling}

We have just outlined how one way to understand the dilaton
shifts appearing in decomposition is through the presence and properties
of interfaces linking the different universes.
In this section we shall outline
another way to understand the dilaton shift 
conjecture~(\ref{eq:dilaton-shift-conj})
in the case of orbifolds and gauge theories.
This alternative understanding uses the fact that
such theories (with 1-form symmetries) can be
represented formally as
theories coupled to a topological field theory, 
specifically,
two-dimensional Dijkgraaf-Witten theory -- the prototypical
example of an orbifold with a trivially-acting subgroup.

Now, to be clear, one should be careful when talking about coupling
physical theories to topological field theories.
Although TFT's are fantastically useful for mathematics applications,
as physical theories they violate basic axioms of field theory
such as unitarity and spin-statistics.  As a result,
one expects that coupling a physical theory to
a TFT ordinarily would ordinarily not not yield a perfectly
well-behaved physical theory -- or
at least, the resulting theory may violate some of the standard
axioms of quantum field theory.  As a result, given any theory that
is described as the result of coupling to a TFT, it behooves one to
check to understand which axioms are violated.
In the present circumstances, two-dimensional Dijkgraaf-Witten theory
is a unitary theory, so unitarity is unbroken, but cluster decomposition
is violated, which ultimately is one way of thinking about the existence
of a decomposition.

Let us make this intuition more precise.  
We compare the partition function of
the general conjecture~(\ref{eq:dilaton-shift-conj}) on a Riemann surface
of genus $g$,
namely
\begin{equation} 
Z_g\left( [X/\Gamma] \right)
 \: = \: |K|^{2g-2} \sum_U (\dim R_U)^{2-2g} Z_g\left( X_U \right),
\end{equation}
to the analogous partition function of a two-dimensional Dijkgraaf-Witten
theory with orbifold group $K$ and twisting $\omega \in H^2(K,U(1))$
on the same Riemann surface,
namely
\begin{equation}
Z_{DW, g} \: = \: 
 |K|^{2g-2} 
\sum_R \left( \dim R \right)^{2-2g}.
\end{equation}
In the Dijkgraaf-Witten partition function $Z_{DW,g}$, the sum is over
all irreducible representations of $K$.  

Comparing the Dijkgraaf-Witten partition function to that of the
general conjecture, they clearly have basic parallels -- both involve
a sum over irreducible representations $R$ of $K$, 
both involve factors of
\begin{equation}
\left( \frac{\dim R}{|K|} \right)^{2-2g}.
\end{equation}

To construct the partition function of the general conjecture from
that of Dijkgraaf-Witten, we restrict the sum over irreducible
representations to a subset (the orbits of a group action on $\hat{K}$),
and we multiply the contribution to each Dijkgraaf-Witten
universe by a factor of the
partition function $Z_g(X_R)$ of a coupled theory.

In passing, note that dilaton shifts in other topological field theories
should be understood a bit differently.  For example, in the $G/G$ model,
the universes are indexed by {\it integrable} 
irreducible representations of $G$.

\section{Dilaton shifts versus probability measures}
\label{sect:ds-vs-prob}

Dilaton shifts often arise in ways that can be interpreted
in terms of probability densities.  We have already
discussed their appearance in asymptotic densities of states
in \cite{Kapec:2019ecr} in section~\ref{sect:consistency-asymp}.
Another simple example arises
in two-dimensional Dijkgraaf-Witten theory.  There, the
dilaton shift factors are (see e.g.~section~\ref{sect:2ddw:1})
\begin{equation}
\left( \frac{\dim R}{|G|} \right)^{2-2g},
\end{equation}
which are clearly related to the Plancherel measure on the set of irreducible
representations of a finite group $G$, a normalized probability density
on the set of irreducible representations whose value for any irreducible
representation $R$ is
\begin{equation}
\frac{ (\dim R)^2 }{|G|},
\end{equation}
see e.g.~\cite{borodin1,Chattopadhyay:2019pkl}.
Related ideas also appear in \cite{Betzios:2022oef}, which discusses superselection sectors
associated to irreducible representatiosn of a gauge group, and
probability densities related to the square of the dimensions of
those representations.

All that said, however, decompositions are not the same as ensembles.

In this section we will do a more careful comparison of the two notions.
We begin in section~\ref{sect:decomp-neq-ensemble} by showing how to
distinguish the two, by working on a spacetime with multiple
connected components.  We further pursue that difference in
section~\ref{sect:syk}, in a comparison of fields appearing in
mirrors to decompositions, compared to stochastic variables appearing in
e.g.~the SYK construction.  The notions appear to be more closely related
on connected spacetimes, and later in section~\ref{sect:ee} we observe
how, at least on connected spacetimes, dilaton shift factors have
been interpreted as probability densities in some entanglement
entropy computations.  We also discuss a generalization of those
entanglement entropy computations.

\subsection{Fundamental distinction}
\label{sect:decomp-neq-ensemble}

It is tempting to relate dilaton shifts to some sort of
probability measure, interpreting universes as events in a probabilistic
ensemble over a space of couplings.
Let $R$ index universes,
and for any one spacetime $X$ over which a quantum field theory is
defined, let $\rho_X(R)$ denote the dilaton shift associated with
universe $R$.  At least on a {\it connected} spacetime,
correlation functions in the theory then have the form
\begin{equation}
\langle {\cal O}_1 \cdots {\cal O}_n \rangle \: = \:
\sum_R \rho_X(R) \langle {\cal O}_1 \cdots {\cal O}_n \rangle_R,
\end{equation}
where $\langle \cdots \rangle_R$ denotes a correlation function in
universe $R$ without a dilaton shift.

However, there is a fundamental difficulty in such an interpretation
which manifests itself on disconnected spacetimes.  Briefly,
\begin{itemize}
\item in a decomposition, one sums over all universes on each
component of the spacetime, as one has a disjoint union of QFTs,
\item whereas in an ensemble, there is one fixed sum over the different
events (universes), independent of the number of components of spacetime.
\end{itemize}
This distinction is visible in e.g.~partition functions.

To be concrete, consider an example of a QFT that decomposes into
$n$ separate universes, schematically
\begin{equation}
X \: = \: Y_1 \coprod Y_2 \coprod \cdots \coprod Y_n.
\end{equation}
Suppose further that our spacetime $\Sigma$ also decomposes into
two pieces,
\begin{equation}
\Sigma \: = \: \Sigma_1 \coprod \Sigma_2.
\end{equation}
Then, in a decomposition, the partition function of theory $X$
on spacetime $\Sigma$ has the form
\begin{equation}
Z(\Sigma, X) \: = \: \left(
\sum_{i=1}^n Z(\Sigma_1, Y_i) \right)
\left(
\sum_{i=1}^n Z(\Sigma_2, Y_i) \right),
\end{equation}
whereas if the decomposition was interpreted as a probabilistic
ensemble, with a probability distribution determined by the
dilaton shift, then
the partition function would be
\begin{equation}
Z(\Sigma, X) \: = \: \sum_{i=1}^n Z(\Sigma_1, Y_i)
Z(\Sigma_2, Y_i).
\end{equation}
In short, the difference between the two interpretations is visible
in the (non)existence of cross terms in the partition function.

\subsection{Decomposition mirror fields versus SYK stochastic parameters}
\label{sect:syk}

To try to further illuminate the difference, we will compare a couple of
closely related examples, namely,
\begin{itemize}
\item the mirror to a GLSM for a gerbe, which admits a locally constant
field valued in roots of unity,
\item Landau-Ginzburg models with a stochastic parameter as in the
SYK model.
\end{itemize}
As before, differences arise over spacetimes with multiple components.

First, we discuss mirrors to two-dimensional
gauged linear sigma models in which a subgroup of the gauge group
acts trivially
(technically, GLSMs for gerbes).
Examples are discussed in \cite{Pantev:2005rh,Gu:2018fpm}.
Briefly, when one computes the mirror to such a theory,
following the usual prescriptions of \cite{Hori:2000kt,Gu:2018fpm},
the mirrors all take the form of Landau-Ginzburg models with
locally-constant
`finite-valued fields.' 
An example of 
the superpotential in these Landau-Ginzburg models has the
form
\begin{equation}
W \: = \: x_1^3 \: + \: x_2^3 \: + \: x_3^3 \: + \: \Upsilon x_1 x_2 x_3
\end{equation}
where $x_{1-3}$ are ordinary chiral superfields, and $\Upsilon$ is
locally constant,
valued in $k$th roots of unity, so that $\Upsilon^k = 1$.  

The path integral's sum over values of $\Upsilon$ is the
sum over universes, with different values of $\Upsilon$ corresponding
to different universes, each of which is, individually, an ordinary
Landau-Ginzburg model, with a different complex structure.

The intuition for this result is straightforward.  The original theory
decomposed, into a disjoint sum 
of theories with different theta angles / $B$ fields.
The mirror to such a decomposition is a disjoint sum of theories with
different complex structures, which is precisely the effect of
the $\Upsilon$ term in the superpotential above.

Because $\Upsilon$ is locally constant, if the spacetime is not connected,
it can take different values on different components, and so on a disconnected
space, the partition function is a product of sums -- we get cross terms,
exactly as described earlier in section~\ref{sect:decomp-neq-ensemble}.

Now, naively, at least on a connected spacetime,
the field $\Upsilon$ appears to be a finite version of a stochastic
variable as has been utilized in the SYK model 
(see e.g.~\cite{Sachdev:1992fk,kitaev}, and \cite{Chang:2023gow} for a GLSM 
version)), in the sense that when one
computes correlation functions, the theory sums over its values,
weighted by probability densities (corresponding to dilaton shifts).
The difference is that in the SYK model, there is one sum, independent
of the number of connected components of spacetime.

For completeness, let us pursue this a bit further, to describe
Landau-Ginzburg models with stochastic variables, as in 
the SYK model.

Consider an ensemble indexed by $\psi \in {\mathbb C}$,
over Landau-Ginzburg models with
superpotential of the
form 
\begin{equation}
W \: = \: x_1^3 \: + \: x_2^3 \: + \: x_3^3 \: + \:
\psi x_1 x_2 x_3,
\end{equation}
and the stochastic variable $\psi$ is weighted by
probability 
$\rho(\psi)$.  Since $\psi$ is not a field, varying over the worldsheet,
but instead an index for universes, the path integral contains only a
single
ordinary integral over its values (independent of the number of
connected components of spacetime).

To be more specific, consider a B-twisted Landau-Ginzburg model with
such a field.  Assume further for simplicity that the vacua are isolated,
and the worldsheet is connected, of genus $g$,
then, correlation functions have the schematic form\footnote{
This is a trivial extension of results in
\cite{Vafa:1990mu}.
}
\begin{eqnarray}
\langle {\cal O}_1 \cdots {\cal O}_n \psi \cdots \psi \rangle
& = &
\int d \psi \, \rho(\psi) \int [D \phi] \exp(-S)  {\cal O}_1 \cdots {\cal O}_n \psi \cdots \psi ,
\\
& = &
\int d \psi \, \rho(\psi) \sum_{dW = 0 }  {\cal O}_1 \cdots {\cal O}_n \psi \cdots \psi
H^{g-1},
\end{eqnarray}
where $H$ is the determinant of the matrix of second derivatives
$\partial_i \partial_j W$, evaluated at critical loci.

On a connected worldsheet, the mirror to a decomposition, described by
a finite-valued $\Upsilon$, has correlation functions computed in
essentially the same form, with $\rho(\Upsilon) = 1$.  On a disconnected
worldsheet, a correlation function is a product of correlation functions
on each component, with a different $\Upsilon$ on each component, instead
of a single overall $\Upsilon$ as would happen in an ensemble.

\subsection{Entropy and dilaton shifts}
\label{sect:ee}

In computations of entropy, dilaton shift between
universes are sometimes interpreted as probability measures --
with all the caveats discussed earlier in this section.
In this section we will briefly describe how various entropy computations
appearing in the literature can be understood in terms of dilaton shifts.

Entanglement entropies have been computed in many references.
Our discussion below will follow the framework of
\cite{Donnelly:2014gva,Donnelly:2016jet,Donnelly:2018ppr,Donnelly:2020teo},
and more specifically,
 \cite[section 1.4]{Donnelly:2014gva}.
(A handful of additional references include
\cite{Callan:1994py,Lewkowycz:2013nqa,Nishioka:2013haa,Nishioka:2016guu,Hubeny:2019bje}, and we emphasize
that the literature contains numerous others.)  Since various entropies
have been discussed in detail in the literature, we shall just summarize
pertinent computations, highlighting the specific relation to dilaton
shifts and decompositions.

Begin with a sphere $S^2$.  Slice it along $n \geq 1$ intervals,
and let $\pm$ denote either side of the cut.
Take $q > 1$ copies of this cut sphere, and let $(i,\pm)$ denote either
side of the cut on the $i$th copy (independently of the choice of interval,
which will all be treated in parallel).  Glue $(i,-)$ to $(i+1,+)$
(and permuted cycically).
Label the result $X^{(q)}$.

Geometrically, $X^{(q)}$ is a branched $q$-fold cover of
$S^2$, branched over $2n$ points (the endpoints of the $n$
intervals), at each of which the branching is maximal (all sheets of the
cover participate).  
From the Riemann-Hurwitz theorem,
\begin{equation}
\chi(X^{(q)}) \: = \: q \chi({\mathbb P}^1) \: - \:
\sum_{i=1}^{2n} (q-1)
\: = \: 2 q - 2n(q-1),
\end{equation}
Geometrically, $X^{(q)}$ is a curve of
genus 
\begin{equation}
g \: = \: (1-n)(1-q).
\end{equation}
The total area of $X^{(q)}$ is $q A$
for $A$ the area of a single $S^2$.

Let $Z(q)$ denote the partition function of a theory on $X^{(q)}$.
Following e.g.~\cite{Donnelly:2014gva}, define the density matrix $\rho$ by
\begin{equation}
{\rm Tr}\, \rho^{q} \: = \: \frac{ Z(q) }{ Z(1)^q },
\end{equation}
and then the replica trick yields the von Neumann entropy
as a limit of the R\'enyi entropy
\cite[equ'n (31)]{Donnelly:2014gva}
\begin{equation}
S 
\: = \: \lim_{q \rightarrow 1} \frac{1}{1-q} \ln  \frac{ Z(q) }{ Z(1)^q }
\: = \:
- \left. \frac{\partial}{\partial q} \frac{ Z(q) }{ Z(1)^{q} }
\right|_{q=1}.   \label{eq:entropy:sq}
\end{equation}

Now, let us apply this to a theory which decomposes.
Write the partition function on a connected spacetime $\Sigma$ in the form
\begin{equation}
Z(\Sigma) \: = \: \sum_R f(R)^{\chi(\Sigma)} Z_R(\Sigma),
\end{equation}
where $R$ indexes universes, $Z_R$ is the partition function for the
theory in universe $R$, and $f(R)$ encodes the dilaton shift.  
Then, in the present case,
\begin{equation}
Z(q) \: = \: \sum_R f(R)^{2q - 2nq + 2n} Z_R(q).
\end{equation}
Define
\begin{equation}
p(R) \: = \: \frac{ f(R)^2 Z_R(1) }{ Z(1) },
\end{equation}
so that
\begin{equation}
\sum_R p(R) \: = \: 1.
\end{equation}
(In effect, $p(R)$ acts analogously to a probability, though in the
spirit of this paper we can also interpret it in terms of dilaton shifts.)

Then, plugging into~(\ref{eq:entropy:sq}),
it is straightforward to compute, at least formally,
\begin{eqnarray}
S & = &
- \left. \frac{\partial}{\partial q} \frac{ Z(q) }{ Z(1)^{q} }
\right|_{q=1},
\\
& = &
\sum_R p(R) \left( - \ln p(R)  +  S_R  + 2 n \ln f(R) \right),
\end{eqnarray}
where $S_R$ represents the result from just universe $R$:
\begin{eqnarray}
S_R & = & - \left. \frac{\partial}{\partial q} \frac{ Z_R(q) }{ Z_R(1)^{q} }
\right|_{q=1},
\\
& = &
\ln Z_R(1) \: - \: \frac{ Z_R'(1) }{ Z_R(1) }.
\end{eqnarray}
The reader should note that
\begin{itemize}
\item the result has the form of a sum over universes,
\item the probability $p(R)$ is proportional to the genus-zero dilaton shift.
\end{itemize}

This result is also closely related\footnote{
We would like to thank O.~Parrikar for pointing this out to us.
}
to an analogous result for 
entropy in the presence of
superselection sectors, where the entropy can be described as a sum of a 
contributions from separate sectors plus a Shannon contribution arising from
just the probabilitiy densities,
see e.g.~\cite[section 3.2, equ'ns (27)-(28)]{Casini:2013rba}, 
\cite{Casini:2019kex,bw}.  The good reason for this relationship is that
in deep IR / infinite volume limits,
superselection becomes decomposition.  The difference is that at finite
energies and finite volumes, in a decomposition one still has a disjoint
union of quantum field theories, which is not true of superselection sectors.
Since superselection has a limit in which it becomes decomposition,
it is natural to expect entropy formulas, for example, to have a similar
form, as we have observed here.
(In passing, see also \cite{Casini:2021tax} for a discussion in terms of 
higher-form symmetries.)

In the special case of a decomposition to invertible field theories,
where
\begin{equation}
Z_R(q) \: = \: \exp\left(-q A f_2(R) \right),
\end{equation}
it is straightforward to check that $S_R = 0$, so that
\begin{equation}
S \: = \: \sum_R p(R) \left( - \ln p(R)   + 2 n \ln f(R) \right).
\end{equation}

Now, let us compare to particular cases.
\begin{itemize}
\item Two-dimensional pure Yang-Mills.  This is discussed in
e.g.~\cite{Donnelly:2014gva,Donnelly:2016jet,Donnelly:2018ppr,Donnelly:2020teo}.
Here, the universes are indexed by irreducible representations $R$ of
the gauge group, and
\begin{equation}
f(R) \: = \: \dim R, \: \: \: f_2(R) \: = \: C_2(R).
\end{equation}
The expression for the entropy,
\begin{equation}
S \: = \: \sum_R p(R) \left( - \ln p(R)  + 2 n \ln \dim R \right),
\end{equation}
matches e.g.~\cite[equ'n (37)]{Donnelly:2014gva}.
\item Nonabelian $BF$ theory.  This is just the zero-area limit of
two-dimensional pure Yang-Mills (see e.g.~\cite[section 2]{Witten:1991we}), 
and so results for
entropy follow immediately from those above.
\item Two-dimensional Dijkgraaf-Witten theory.  Here,
the universes are indexed by irreducible (projective) representations
$R$ of the orbifold group $G$, and
\begin{equation}
f(R) \: = \: \frac{\dim R}{|G|}, \: \: \: f_2(R) = 0,
\end{equation}
so
\begin{equation}
p(R) \: = \: f(R)^2 \: = \: \left(  \frac{\dim R}{|G|} \right)^2,
\: \: \:
Z_R(q) \: = \: \exp(-q A f_2(R) ) \: = \: 1,
\end{equation}
and
\begin{equation}
S \: = \: \sum_R p(R) \left( - \ln p(R) \: + \: 2 \pi \ln\left(
 \frac{\dim R}{|G|} \right) \right).
\end{equation}
\item Two-dimensional unitary topological field theories.
In a modification of the notation of section~\ref{sect:tft:genl},
if we index constituent universes by $R$ to write the partition function of
a two-dimensional topological field theory on a Riemann surface of genus
$g$ as~(\ref{eq:2dtft:partfn})
\begin{equation}
\sum_R \left( \theta_R \right)^{1-g},
\end{equation}
then we take $f(R) = \sqrt{ \theta_R} \propto \sqrt{\dim_q \Pi_R}$, 
$f_2(R) = 0$,
so that
\begin{equation}
p(R) \: = \: \frac{ \theta_R }{ Z(1) }.
\end{equation}
\end{itemize}

\section{Conclusions}

In this paper we have studied dilaton shifts (Euler counterterms)
weighting the different universes of decompositions in two-dimensional
quantum field theories.  Although these shifts are just counterterms,
they arise in a more or less canonical form, determined by the dimension
of representations indexing the universes, whose form we have discussed
in detail, and as is expected from global symmetries.
We have outlined consequences for volumes of moduli spaces of flat connections,
and also discussed distinctions with and relations to probability measures
in several contexts.

\section{Acknowledgements}

We would like to thank F.~Benini, S.~Datta,
D.~Freed, S.~Gukov, S.~Hellerman, L.~Jeffrey, L.~Lin, O.~Parrikar,
A.~Perez-Lona,
D.~Robbins, Y.~Tachikawa,
M.~\"Unsal, T.~Vandermeulen, and X.~Yu
for useful discussions.
E.S. was partially supported by NSF grant
PHY-2310588.

\appendix

\section{Review of invertible field theories}
\label{app:inv}

The notion of invertible field theories arise when discussing tensor
products of quantum field theories.  In a product $A \otimes B$
of quantum field theories $A$, $B$, (distinguished
from a disjoint union or sum that play a role in decomposition),
the Fock space is a tensor product of the Fock spaces of $A$ and $B$ separately,
and the partition function of $A \otimes B$ is a product of the
partition functions of $A$ and $B$ separately.

An invertible field theory is a quantum field theory that is invertible
under such a product operation, which implies, for example, that its Fock
space is one-dimensional -- the only states are scalar multiples of the vacuum.

A prototype for an invertible field theory in two dimensions
is a sigma model whose target space
is a single point, with vanishing action.  Given such a sigma model,
we can still add counterterms.  For example, we can consider a two-parameter
family of counterterms described by the action
\begin{equation}
S \: = \: \int_{\Sigma} d^2 x \, \sqrt{g} \left( \lambda_1 \: + \:
\lambda_2 \frac{R}{4 \pi} \right),
\end{equation}
where $g$ is the (classical, nondynamical) metric on the 
worldsheet $\Sigma$, and $R$ is the Ricci sclar of $\Sigma$, so that the
partition function at genus $g$ is
\begin{equation}
Z \: = \: \exp\left( \lambda_1 (\mbox{area of }\Sigma)
 \: + \: \lambda_2 \chi(\Sigma)
\right),
\end{equation}
where $\chi(\Sigma)$ is the Euler characteristic of $\Sigma$.

Elsewhere in this paper, we denote this family of invertible field theories
by Inv$(\lambda_1,\lambda_2)$.

\section{Finite group representation theory identities}
\label{app:reptheory}

In this appendix we collect some identities arising in 
the representation theory of a finite group $G$,
twisted by a cocycle $[\omega] \in H^2(G,U(1))$,
normalized so that $\omega(1,g) = \omega(g,1) = 1$.
These identities can be found in
\cite[appendix B]{Ramgoolam:2022xfk},
\cite[appendix B]{Sharpe:2021srf}, and
references therein:
\begin{equation} \label{eq:char-om1}
\frac{1}{|G|} \sum_{g \in G} 
\frac{ \omega(a,g) \omega(g^{-1},b) }{ \omega(g,g^{-1}) } \,
\chi_R(ag) \, \chi_S(g^{-1} b)
\: = \:
\frac{ \delta_{R,S} }{\dim R} \, \omega(a,b) \, \chi_R(ab).
\end{equation}
\begin{equation} \label{eq:char-om1r}
\frac{1}{|G|} \sum_{g \in G} 
\frac{ \omega(g,a) \omega(b,g^{-1}) }{ \omega(g,g^{-1}) }
\chi_R(ga) \chi_S(bg^{-1})
\: = \: 
\frac{\delta_{R,S}}{\dim R} \omega(a,b) \chi_R(ab).
\end{equation} 
\begin{equation} \label{eq:char-om2}
\frac{1}{|G|} \sum_{g \in G} \frac{ \omega(g,a) \, \omega(g^{-1},b) \,
\omega(ga, g^{-1} b) }{ \omega(g,g^{-1}) } \,
\chi_R(g a g^{-1} b) \: = \:
\frac{1}{\dim R} \, \chi^R(a) \, \chi_R(b).
\end{equation}
\begin{equation}  \label{eq:char-om2r}
\frac{1}{|G|} \sum_{g \in G} \frac{
\omega(a,g) \, \omega(b,g^{-1}) \, \omega(ag,bg^{-1}) }{ \omega(g,g^{-1}) }
\chi^R(a g b g^{-1}) \: = \: 
\frac{1}{\dim R} \, \chi_R(a) \chi_R(b).
\end{equation}
and also, for $[g]$, $[h]$ both\footnote{
If either is not an $\omega$-regular conjugacy class, then the corresponding
characters vanish, and the sum equals zero.
} $\omega$-regular conjugacy classes,
\begin{eqnarray} 
\sum_R \frac{
\chi_R(g) \chi_R(h^{-1})
}{
\omega(h,h^{-1})
}
& = & \left\{ \begin{array}{cl}
0 & g, h \mbox{ not conjugate}, \\
\frac{ |G|}{|[g]|}  & g = h, \\
\frac{ \omega(a,g) }{ \omega(h,a) } \frac{ |G|}{|[g]|}
 & g = a^{-1} h a,
\end{array} \right.
\label{eq:char-master2}
\end{eqnarray}
where $R$, $S$ are irreducible projective representations
(with respect to $\omega$).

\section{Two-dimensional Dijkgraaf-Witten theory, from triangulations}
\label{app:2ddw:triangle}

It is a standard result that two-dimensional pure Yang-Mills theory can
be described by associating data to a triangulation of a Riemann surface,
and then gluing along edges.  In this appendix we will review the 
analogous construction for two-dimensional Dijkgraaf-Witten theory.
(See also e.g.~\cite[appendix A]{Gardiner:2020vjp} 
for a related state sum construction.)

Specifically, we will describe
computations in Dijkgraaf-Witten theory
in terms of data assigned to cylinders, disks, and pairs-of-pants,
and how they are glued, close to the spirit of pure Yang-Mills.

Recall that the partition function\footnote{
This is closely related to, but slightly different from, results
for Dijkgraaf-Witten correlation functions.  Specifically, a correlation
function has the form (see e.g.~\cite{Ramgoolam:2022xfk})
\begin{equation}
\langle U_1 \cdots U_b \rangle \: = \:
\sum_R \left( \frac{\dim R}{|G|} \right)^{\chi(\Sigma)}
\frac{\chi_R(U_1)}{\dim R} \cdots \frac{\chi_R(U_b)}{\dim R},
\end{equation}
which differs from the result above for partition functions with
boundary components by factors of $\dim R$, reflecting differences
in the normalizations of states.
The different factors of $\dim R$ in the boundary case make the
gluing construction possible.
} of a two-dimensional (possibly twisted)
Dijkgraaf-Witten theory with $b$ boundary components has the form
\begin{equation}
Z(\Sigma, U_1, \cdots, U_b) \: = \:
\sum_R \left( \frac{\dim R}{|G|} \right)^{\chi(\Sigma)}
\chi_R(U_1) \cdots \chi_R(U_b),
\end{equation}
where the sum is over irreducible projective representations $R$ of $G$,
twisted by the cocycle $[\omega] \in H^2(G,U(1))$
(itself normalized by $\omega(1,g) = \omega(g,1) = 1$).
This can be described axiomatically in the same sense as two-dimensional
pure Yang-Mills,
with gluing accomplished via
\begin{equation}
\frac{1}{|G|} \sum_{U \in G} \frac{1}{\omega(U,U^{-1})}.
\end{equation}

Listed below are the partition functions for some standard examples:
\begin{enumerate}
\item Disk:
\begin{equation}
Z_{\rm disk}(U) \: = \: \sum_R \left( \frac{\dim R}{|G|} \right) \chi_R(U).
\end{equation}
\item Cylinder:
\begin{equation}
Z_{\rm cylinder}(U_1,U_2) \: = \:
\sum_R \chi_R(U_1) \chi_R(U_2^{-1}).
\end{equation}
\item Pair of pants:
\begin{equation}
Z_{\rm pants}(U_1,U_2,U_3) \: = \: 
\sum_R \left( \frac{\dim R}{|G|} \right)^{-1}
\chi_R(U_1) \chi_R(U_2) \chi_R(U_3).
\end{equation}
\end{enumerate}

The key identity needed to implement the gluing is
\begin{equation}
\frac{1}{|G|} \sum_{U \in G} \frac{1}{\omega(U,U^{-1})}
\chi_R(U) \chi_S(U^{-1}) \: = \: \delta_{R,S}.
\end{equation}
Further identities of this form are given in appendix~\ref{app:reptheory}.

As a consistency check, let us formally glue a cylinder to a disk:
\begin{eqnarray}
\lefteqn{
\frac{1}{|G|} \sum_{U \in G} \frac{1}{\omega(U,U^{-1})} \left[
\sum_R \chi_R(U_1) \chi_R(U^{-1}) \right]
\left[ \sum_S \left( \frac{\dim S}{|G|} \right) \chi_S(U) \right]
} \nonumber \\
& = &
\frac{1}{|G|} \sum_{R,S} \chi_R(U_1) \left( \frac{\dim S}{|G|} \right) 
|G| \delta_{R,S},
\\
& = &
\sum_R  \left( \frac{\dim R}{|G|} \right) \chi_R(U_1),
\end{eqnarray}
which is precisely the partition function of a disk, as expected.

\end{document}